\documentclass[11pt]{article} 
\usepackage[utf8]{inputenc}
\usepackage{amsmath}
\usepackage{amssymb}
\usepackage{comment}
\usepackage{float}

\usepackage{caption}
\usepackage{xcolor}
\definecolor{dgreen}{rgb}{0,0.5,0}
\definecolor{dpink}{rgb}{1,0.3,0.3}
\definecolor{darkblue}{rgb}{0,0,0.6}
\definecolor{purple}{rgb}{0.4,.2,0.7}
\usepackage[margin = 2.5cm]{geometry}
\usepackage{graphicx}
\usepackage[hyperfootnotes = false, colorlinks = true, linkcolor = blue, citecolor = purple]{hyperref}
\usepackage{subcaption}

\usepackage{ytableau}

\usepackage{fancyhdr}
\parskip=\medskipamount
\newgeometry{margin=2cm}

\def\la{\label}
\def\nref#1{(\ref{#1})}

	\newcommand{\hf}{\frac{1}{2}}

	\newcommand*\diff{\mathop{}\!\mathrm{d}}
	
	\usepackage{physics}
	\renewcommand{\ev}[1]{\langle #1 \rangle} %
	
	\def\i{\mathrm{i}}

\def\red{\color{red}}
\usepackage{tikz}
\usepackage{makecell}

\begin{document}
\renewcommand*{\thefootnote}{\fnsymbol{footnote}}

\thispagestyle{empty}
\begin{center}
    ~\vspace{5mm}

  {\LARGE \bf Bootstrapping Ground State Correlators\\ \vspace{0.25cm} in Matrix Theory, Part I}

   \vspace{0.5in}
\begingroup\large
   {\bf Henry W. Lin\footnote{hlin2@stanford.edu} $\,$ \it{and} $\,$ \bf Zechuan Zheng\footnote{zechuan.zheng.phy@gmail.com}}
\par\endgroup
    \vspace{0.5in}

\begingroup\sffamily\footnotesize
    \textsuperscript{$\star$}Stanford Institute for Theoretical Physics, Stanford University, Stanford, CA 94305\\
   \textsuperscript{$\dagger$}Perimeter Institute for Theoretical Physics, Waterloo, ON N2L 2Y5, Canada
 \par\endgroup               
    \vspace{0.5in}
    
\end{center}

\vspace{0.5in}

\begin{abstract}
The D0-brane/Banks-Fischler-Shenker-Susskind matrix theory is a strongly coupled quantum system with an interesting gravity dual.
We develop a scheme to derive bootstrap bounds on simple correlators in the matrix theory at infinite $N$ at zero energy by imposing the supercharge equations of motion. By exploiting SO(9) symmetry, we are able to consider single-trace operators involving words of length up to 9 using very modest computational resources. We interpret our initial results as strong evidence that the bootstrap method can efficiently access  physics in the strongly coupled, infinite $N$ regime. %

\end{abstract}

\vspace{0.5cm}

\thispagestyle{empty}

\newpage

\setcounter{tocdepth}{3}

\tableofcontents
\renewcommand*{\thefootnote}{\arabic{footnote}}
\setcounter{footnote}{0}
\section{Introduction}

Strongly coupled quantum systems with a large number of degrees of freedom hold many secrets. An interesting example is the Banks-Fischler-Shenker-Susskind (BFSS) matrix theory \cite{deWit:1988wri, Banks:1996vh} or D0-brane quantum mechanics in the large-$N$ limit. According to \cite{Banks:1996vh, Itzhaki:1998dd}, solving the strongly coupled dynamics in the BFSS model is tantamount to solving $M$-theory/type IIA string theory\footnote{with certain asymptotic boundary conditions.}. This theory is perhaps the simplest example of gauge/gravity duality in that the gauge theory is merely a $0+1$ dimensional quantum system as opposed to a quantum field theory (for a recent review, see \cite{Maldacena:2023acv}).

Solving such a model is a longtime dream. By computing observables in the matrix theory as a function of various parameters, one could learn a lot about quantum gravity beyond semiclassical Einstein gravity. Indeed, heroic lattice Monte Carlo simulations \cite{Kabat:2000zv, Anagnostopoulos:2007fw, Hanada:2008ez, Catterall:2008yz, Filev:2015hia, Kadoh:2015mka, Berkowitz:2016jlq, Berkowitz:2018qhn, Pateloudis:2022ijr} have already led to a non-trivial prediction about $(\alpha')^3$ corrections to type IIA supergravity that has not yet been verified by any other computational technique in string theory.

Applying the Monte Carlo method to this system is a highly non-trivial task. First, large-$N$ extrapolation is computationally expensive, especially when the number of lattice sites is large, as required for accessing low temperatures where the model is strongly coupled. Second, at finite $N$, the model is only metastable due to the presence of flat directions. 
Third, Monte Carlo is best suited for problems in Euclidean signature where the measure is positive, but integrating out the fermions in the BFSS matrix model yields a measure for the bosons that is complex\footnote{See however \cite{Catterall:2008yz, Catterall:2009xn, Berkowitz:2016jlq} for a study of the sign problem. They argue that for some range of parameters one may simply replace the complex measure with its absolute value.}.

In this work, we will explore a less mature alternative to Monte Carlo, sometimes referred to as the ``matrix bootstrap'', developed in \cite{Anderson:2016rcw,Lin:2020mme,Han:2020bkb,Kazakov:2021lel,Kazakov:2022xuh, Cho:2022lcj,Lin:2023owt, Kazakov:2024ool, Li:2024wrd}. This method is promising as it can directly access the zero energy sector, which is complementary to Monte Carlo (for which low energies/large Euclidean times are the most challenging to access numerically).
Consider a supersymmetric quantum system with a ground state $\ket{\Omega}$ that is annihilated by one (or more) supercharges: $Q \ket{\Omega} = 0$.  Conceptually, the bootstrap method starts by ``guessing'' the expectation values of some set of ``simple'' operators $\ev{\mathcal{O}_i} = \mel{\Omega}{\mathcal{O}_i}{\Omega}$. Then one generates more expectation values $\ev{\mathcal{O}'_j}$ by using the property that for any\footnote{This equation is true for both fermionic and bosonic operators, but in the large-$N$ limit it is only useful for fermionic operators. For bosonic operators, one should instead consider the equation $\ev{ [Q, \mathcal{O}_B] } = 0$. However, for the BFSS model, the supercharge is an SO(9) spinor, so SO(9) symmetry already enforces the commutator condition. } operator $\mathcal{O}$, $\mel{\Omega}{ \{ Q, \mathcal{O}\}  }{\Omega}  = 0$.
To check the initial guess, we construct an operator $\mathcal{V}$ that is a superposition of all the operators which we know something about. If the initial guess is correct, positivity of the Hilbert space inner product guarantees:
\begin{align}
\label{basic}
\langle \mathcal{V}^\dagger \mathcal{V} \rangle \ge 0.
\end{align}
So if we manage to find a $\mathcal{V}$ such that violates \nref{basic}, we rule out our initial guess. By searching over possible guesses (which can be done efficiently using semi-definite programming), one obtains an allowed region in the space of correlation functions. 

This bootstrap approach has the advantage that one can work directly in the infinite $N$, 't Hooft limit by imposing large-$N$ factorization on the correlation functions.
(In Matrix theory, the observables which satisfy large-$N$ factorization more directly probe the type IIA limit \cite{Itzhaki:1998dd} as opposed to the M-theory limit. See section \ref{sec: discuss} and \cite{Polchinski:1999br} for more discussion.) Furthermore, fermions are not a problem since \eqref{basic} only uses Hilbert space positivity. However, a potential pitfall of the bootstrap approach is that for systems with multiple matrices, there is a proliferation of possible correlation functions that enter the bootstrap problem. For $D$ matrices, the number of single-trace operators of a given length $L$ grows exponentially like $\sim D^L$. In the BFSS model, there are $9$ bosonic matrices and their canonical conjugates, in addition to 16 fermionic matrices. This naively suggests that a bootstrap study of BFSS will be prohibitively expensive, but in this paper we will show that by leveraging all the symmetries of the problem, we can generate non-trivial bounds on simple correlators with only modest computational resources.

A non-systematic bootstrap approach to the BFSS correlators was taken in \cite{Lin:2023owt}. Readers less familiar with the matrix bootstrap might find the simpler calculations in \cite{Lin:2023owt} pedagogical. In an upcoming work \cite{LinZheng2}, we will explain how to fully automate the computation of the group theory factors, which will allow us to derive constraints at higher levels. We will also explore constraints on other operators and discuss bootstrapping other matrix models. %

\subsection{The model and its symmetries}
The BFSS matrix theory consists of 9 bosonic matrices $X_I$ and 16 fermionic matrices $\psi_\alpha$, which transform under an SO(9) $R$-symmetry in the fundamental and spinor representations.  All matrices are taken to be Hermitian and traceless; they satisfy the canonical commutation relations:
\begin{equation}
\begin{split}
\label{canonical}
 \quad [X^I_{ij},P^J_{kl}] = \i \, \delta_{il} \delta_{jk} \delta^{IJ} , \quad \{  \psi_\alpha , \psi_\beta \} =   \delta_{\alpha \beta}  \delta_{il} \delta_{jk}   %
\end{split}
\end{equation}
In this work, there is no distinction between upper and lower indices.
The Hamiltonian is
\begin{align}\la{ham}
H &= \hf  \operatorname{Tr}  \left(g^2   {P_I^2}-\frac{1}{2g^2}\left[X_I, X_J\right]^2-\psi_\alpha \gamma^I_{\alpha \beta} \left[ X_I,\psi_\beta\right] \right) .
\end{align}
In the above expression, there is an implicit sum over $I,J$. 
With these conventions, $X$ has units of energy and $g^2$ has units of $E^{3}$.
We can take the SO(9) gamma matrices $\gamma^I$ to be real, traceless, and symmetric\footnote{See Appendix~\ref{app: GammaAlgebra} for the convention of gamma matrices.}.
This model has 16 supercharges which transform as spinors under the SO(9) global symmetry:
The 16 Hermitian supercharges are
\begin{align} \label{eq:susy}
    Q_\alpha &= g \Tr P_I \gamma^{I}_{\alpha \beta} \psi_\beta - \frac{\i}{2g} \Tr [X^I, X^J]\gamma^{IJ}_{\alpha \beta}\psi_\beta,
\end{align}
They satisfy the supersymmetry algebra
\begin{align}\label{eq: gaugegen}
    \{Q_\alpha, Q_\beta\} &= 2 \delta_{\alpha \beta} H +2 \gamma^I_{\alpha \beta} \Tr X^I C\\
    C &= -\i [X^I, P^I]-\psi_{\alpha} \psi_{\alpha} - N {\bf 1}
\end{align}
Without loss of generality\footnote{If we rescale $X \to g^{2/3} X$ and $P \to g^{-2/3} P$, we find
$H \to   \frac{g^{2/3}}{2} \sum_I \operatorname{Tr}\left( P_I^2-\frac{1}{2 }\sum_{J}\left[X_I, X_J\right]^2-\psi_\alpha \gamma_{\alpha \beta}^I\left[X_I, \psi_\beta\right]\right)$.} we may set $g=1$.
In the above equation, $C_{ij}$ is the generator of SU$(N)$ symmetry, where each matrix transforms in the adjoint representation. By choosing the matrices to transform in SU$(N)$ as opposed to U($N$), we have removed the center of mass degree of freedom.

The model has been argued \cite{Yi:1997eg, Moore:1998et, Konechny:1998vc, Porrati:1997ej, Sethi:2000zf} to have a unique normalizable ground state, which preserves SUSY. It follows that such a state must preserve $\text{SO}(9)$ symmetry, supersymmetry, and discrete symmetries.

\section{Bootstrap ingredients}

\subsection{Variables}\label{sec: variables}
Due to large-$N$ factorization, we only need to consider single-trace operators. The operators themselves are ``words" composed of ``indexed letters" $X^I$, $P^J$, and $\psi_{\alpha}$. 
Since $\Tr \psi_\alpha \psi_\alpha=8 N^2$ and $\Tr X^I P^I = \frac{9}{2} \i N^2$, we will perform the following shift:
\begin{equation}\label{largeNscaling}
    \Tr \rightarrow \tr = \frac{1}{N}\Tr, \quad
    X\rightarrow\frac{1}{\sqrt{N}} X, \quad  P\rightarrow\frac{1}{\sqrt{N}} P,\quad \psi\rightarrow\frac{1}{\sqrt{N}} \psi, 
\end{equation}
With this convention, single-trace operators will have an expectation value of $O(1)$.
Because the ground state is $\text{SO}(9)$ invariant, only SO(9) singlets yield non-vanishing expectation values. So the variables in the bootstrap formulation are:
\begin{align} \la{wordInd}
 &\langle\tr \mathcal{O}  \rangle, \quad  \mathcal{O} =\mathcal{O}_{\texttt{ind}} \,\mathcal{I}^{\texttt{ind}},\\  
\nonumber \text{e.g.} \quad 
  & \ev{ \tr \psi_{\alpha_1} \psi_{\alpha_2} \psi_{\alpha_3} \psi_{\alpha_4}} \gamma^{I}_{\alpha_1 \alpha_2} \gamma^{I}_{\alpha_3 \alpha_4 }, \; \ev{ \tr X^{I_1} \cdots X^{I_9}} \epsilon^{I_1 I_2 \cdots I_9}
\end{align}
Here, $\mathcal{O}_{\texttt{ind}}$ is a word of operators composed of  indexed letters ``indexed letters" $X^I$, $P^J$, and $\psi_{\alpha}$, and $\mathcal{I}^{\texttt{ind}}$ denotes an invariant tensor with the given index $\texttt{ind}$. (More generally, we  consider operators with open indices, e.g., $O_\alpha = O_{\alpha, \texttt{ind}} \mathcal{I}^{\texttt{ind}}$.)\footnote{We are saying that for practical implementation, we treat the invariant tensors as symbols, as opposed to treating each component of the operator $O_\texttt{ind}$ as a variable.}
The algebra of invariant tensors can be generated from
\begin{equation}
        \delta^{IJ}, \quad \epsilon^{I_1 I_2 \ldots I_9}, \quad \gamma^I_{\alpha\beta}.
\end{equation}
Denote the set of indices in \nref{wordInd} as $\{ I_1, I_2, \cdots, I_B, \alpha_1, \alpha_2, \cdots \alpha_F \} $. For fixed $B$ and $F$, each invariant tensor is a linear map 
\begin{align}
   \mathcal{I} :  \overbrace{{\bf 9} \otimes \cdots \otimes {\bf 9}}^{B \text{ vectors}} \otimes  \overbrace{{\bf 16} \otimes \cdots \otimes {\bf 16}}^{F \text{ spinors} }  \to {\bf 1},
\end{align} where ${\bf 9}, {\bf 16}$, and ${\bf 1}$ denotes the vector, spinor, and singlet irreps of SO(9). The number of invariant tensors is precisely the multiplicity of the singlet irrep in the decomposition of the tensor product. We used the \texttt{Mathematica} package \texttt{GAMMA} \cite{Gran:2001yh} for some of the $\gamma$ algebra computations; the package \texttt{LieART} was also useful \cite{Feger:2019tvk}.

In practice, by only working with SO(9) invariants, we dramatically reduce the number of variables in the bootstrap problem. The price to pay is that with more than two indices, there are various ``channels'' corresponding to different choices of decompositions of the tensor product. In general, we must solve the crossing kernel to express all these different decompositions in terms of a standard basis; an example of this is discussed around  \nref{defCrossing}.

\subsection{Kinematic Constraints}
\def\O{\mathcal{O}}
We refer to constraints that do not involve the explicit form of the Hamiltonian or supercharge as "kinematic constraints." They can be divided into three categories. %
\begin{enumerate}
    \item {\bf Cyclicity}: we enforce cyclicity of the trace\footnote{Here we are referring to the trace over matrices that transform under SU($N$), not to be confused with the trace over the quantum Hilbert space}. If all the matrices in a given word commute, this would simply be the condition $\tr \O_1 \O_2 \cdots \O_n = \tr \O_2 \cdots \O_n \O_1 $. However, since each matrix element is really a quantum operator satisfying \nref{canonical}, we must take this into account. In general, we obtain %
    \begin{align} \label{eq: cyc}
        \tr \O_1 \O_2 \cdots \O_n = \pm \tr \O_2 \cdots \O_n \O_1 + \text{double trace} ,
    \end{align}
    where $+$ $(-)$ corresponds to a bosonic (fermionic) operator $\mathcal{O}_1$. For example, using the normalizations in \nref{largeNscaling},
    \begin{align} \label{eq: exampleCyc}
       \ev{ \tr X^{I_1}  X^{I_2} X^{I_3} P^{I_4} X^{I_5} X^{I_6}}  = \ev{\tr   X^{I_2} X^{I_3} P^{I_4} X^{I_5} X^{I_6} X^{I_1}} + \i \ev{\tr X^{I_2} X^{I_3}} \ev{\tr X^{I_5} X^{I_6}} \delta^{I_1 I_4} 
    \end{align}
    A fermionic example is presented in (\ref{4fermion_cyc}). Here we have imposed {\bf  large-$N$ factorization}, by replacing the expectation value of a double trace with the product of single-trace expectation values.  %

    \item {\bf Hermiticity/time reversal}: Each matrix $\O_i \in \{X^I,P^J,\psi_\alpha \}$ is a Hermitian matrix so $\overline{\ev{\tr \O_1  \cdots \O_n}} = \ev{(\O_n)_{i_n,i_1}^\dagger  \cdots (\O_1)_{i_1, i_2}^\dagger } = \ev{(\O_n)_{i_1, i_n} \cdots (\O_1)_{i_2,i_1}} = \ev{\tr (\O_n \cdots \O_1 )}$. Time reversal invariance implies that correlators that contain an even number of $P$'s are real, and those with an odd number of $P$'s are imaginary\footnote{In practice, we consider words that only involve $\mathcal{P} = - \mathrm{i} P$ instead of $P$, so all correlators are real.}. The conclusion is that correlators that are ``reversed'' words are simply related $\ev{\tr O_1 \cdots O_n} = \pm \ev{\tr O_n \cdots O_1}$.

    \item {\bf Gauge invariance}: one can consider the D0-brane quantum mechanics with or without \cite{Maldacena:2018vsr} the gauge constraint. In either case, the ground state of BFSS is a gauge singlet. This implies that 
    \begin{equation} \label{eq: gaugeConstraint}
        \langle \tr  \mathcal{O} \, C \rangle = 0,\quad \forall \mathcal{O},
    \end{equation}
    where $C$ is the gauge generator, defined in Eq.~\eqref{eq: gaugegen}.
\end{enumerate}

\subsection{Dynamical Constraints}
As mentioned in the introduction, we would like to impose the ground state equation $Q_\alpha \ket{\Omega} = 0$. At infinite $N$, this boils down to the condition
\begin{align}
    \langle \{ Q_\alpha , \mathcal{O}_\alpha \} \rangle = 0, \la{groundBoot} %
\end{align}
where $O_\alpha$ is any single-trace SO(9) spinor. This equation is only non-trivial if $O_\alpha$ is an SO(9) spinor, since the addition of angular momentum rules state that ${\bf 16} \otimes R \supseteq  {\bf 1}$ only if $R = {\bf 16}$. This also of course implies that $O_\alpha$ is fermionic, which is why we never consider replacing the anti-commutator with the commutator (at infinite $N$) in equation \nref{groundBoot}. %

Note that the process of enumerating SO(9) spinors is essentially the same as enumerating SO(9) singlets, which we discussed in subsection \ref{sec: variables}. We enumerate all tensors which would give an SO(9) singlet if tensored with an additional SO(9) spinor.

\subsection{Positivity and SO(9) blocks}
\def\RR{\mathcal{R}}
\def\O{\mathcal{O}}
\def\i{\mathbf{i}}
\def\j{\mathbf{j}}

We would like to consider the general positivity constraints on SO(9) singlet operators. However, to derive these positivity constraints, one must consider SO(9) non-singlets in the intermediate steps. As a trivial example, the expectation value $\ev{\tr X^I X^I} \ge 0 $. To prove this, we observe that this operator is the square of the operator $X^I$ which is an SO(9) vector. Thus, even though SO(9) vectors (or other non-singlet irreps) have vanishing expectation values, we must consider operators that transform under all possible irreps of SO(9) in order to derive the full set of positivity constraints on the SO(9) singlets.

Let's explain the general strategy.
Consider an operator with spinor or vector indices, which we collectively denote $\i = \{I_1, I_2, \cdots, I_n, \alpha_1, \cdots \}$. Equation (\ref{basic}) becomes the positivity constraint
\begin{align}
    \mathcal{M}_{\i \j } =  \ev{ \tr \, \bar{O}_\i O_\j } \succeq 0,
\end{align}
where $\succeq$ means that all eigenvalues of the matrix $\mathcal{M}$ are non-negative.
If we write out this matrix explicitly, the total number of possible values for $\i$ will quickly make this positivity matrix intractable. For example, if we consider operators like $O_\i = X^{I_1} X^{I_2} X^{I_3} X^{I_4}$, we have a matrix of size $6561\times 6561$. Positivity of such a matrix would be a tiny subset of the level 8 constraints; with this ``explicit matrix'' approach, we would have a nearly intractable SDP problem at level 8. Our goal instead is to use group theory to boil down all the positivity constraints of this explicit matrix into a much smaller set of positivity constraints on the SO(9) singlets.

To this end, we first decompose the operators into irreps, e.g.,%
\begin{align}
  O_\i = \sum_R  \sum_{r=1}^{\dim R} (C_R)^{r}_\i (O_R)_r 
\end{align}
An irrep appears in the sum multiple times if the decomposition has multiplicity\footnote{An example where this occurs is the decomposition of the operator $O_\i = X^{I_1} X^{I_2} X^{I_3}$ into irreps: 
\begin{align}
    {\bf 9} \times \mathbf{9} \times \mathbf{9} = 3 (\mathbf{9}) + \mathbf{84} + \mathbf{156} + 2 (\mathbf{231} ). \end{align}
    There are three degenerate vector irreps corresponding to $\{ X^{I} X^{I} X^J, X^{I} X^J X^I, X^J X^I X^I\}$. The fully symmetric and fully anti-symmetric tensors appear with multiplicity one. There are also two  ${\bf 231}$ irreps corresponding to the Young projectors
\ytableausetup{centertableaux}
\def\id{\mathbf{1}}
\begin{align}
    {\bf 231_+ } \; =\; \begin{ytableau}
       1 & 2  \\
  3\\ \end{ytableau} \; , \qquad 
    {\bf 231_-} \; = \; \begin{ytableau}
       1 & 3  \\
  2\\ \end{ytableau} .
\end{align}
} %
Then SO(9) invariance of the state implies that 
\begin{align}
    \ev{\tr (\bar{O}_{\bar{R}})_{\bar{r}} (O_{R})_{r}} = \delta_{\bar{R},R} \delta_{\bar{r},r} a_{\bar{R},R},\\
        \mathcal{M}_{\i \j } = \sum_{R,\bar{R}, r} a_{\bar{R}, R} (C_{\bar R})_\i^r (C_R)_\j^r.
\end{align}
Here the symbol $\delta_{R,R'} = 1$ if $R$ and $R'$ are equivalent representations of SO(9), or else $\delta_{R,R'} = 0$. Thus we have parameterized a large matrix $\mathcal{M}$ in terms of a smaller number of coefficients $a_{\bar{R}, R}$. 
These coefficients are precisely just the SO(9) singlet operators, e.g.,
\begin{align}
    a_{R,\bar{R}} = \ev{\tr (\bar{O}_{\bar{R}})_{r} (O_{R})_{r}}.
\end{align}
Furthermore, we can simplify the positivity requirement $\mathcal{M} \succeq 0$ by evaluating the requirement on a nice basis of vectors $\{ e_A\} $. %
A particularly nice basis is the following. First we view $(C_R)$ as a projector from the vector space indexed by $\mathcal{I}$ to the irrep $R$ (a smaller vector space indexed by $r$). Then we can define the basis $e_A$ to be a collection of the transposed projectors, where $A = (R_A, r_A)$. This spans the bigger vector space of dimension $|\mathcal{I}|$ and satisfies
\begin{align}
  \sum_\j  e_{B}^{\j} (C_R)_\j^r = \delta_{R,R_B} \delta_{r_B}^r.
\end{align}
Then the positivity requirement can be expressed as %
\begin{align}
   \mathcal{M}_{AB} =  \bar{e}_{A}^{\i}  \mathcal{M}_{\i\j} e_{B}^{\j}  =  \sum_{R,\bar{R}} a_{\bar{R}_A, R_B} \delta_{\bar{R}_A,R_B}\delta_{\bar{r}_A, r_B} \succeq 0 . 
\end{align}
The conclusion is that we only need to impose
\begin{align}
    a_{\bar{R}, R } \succeq 0, \quad \bar{R} \sim R.
\end{align}
Note that even if the decomposition of $\i$ contains irrep $R$ with unit multiplicity, the generalization to multiple operators will typically lead to a non-trivial matrix $a$.

\subsection{Hierarchy}\label{sec: hierarchy}
The equality constraints and positivity conditions discussed in this section apply, in principle, to all operators. However, for practical implementation of the bootstrap, we must always make a finite selection. Following the approach of \cite{Kazakov:2024ool}, we introduce a hierarchy among the set of all variables and make our selection based on the level of the variable in this hierarchy. Since any non-vanishing variable always contains an even number of $\psi$ fields, all variables are assigned an integer level.

The hierarchy is defined by sorting operators into {\it levels}: we assign the basic fields $\ell(X) = 1$, $\ell(P) = 2$, and $\ell(\psi) = 3/2$. The level of an operator\footnote{More precisely, for a linear combination of operators, we take the level of the linear combination to be the maximal level of each term. For example, we consider the operator in \eqref{eq: exampleCyc} is a level $7$ operator. The double trace term is a level 4 operator. } is the sum of the levels of all its fields: schematically, $\ell(X^{n_X} P^{n_P} \psi^{n_\psi}) = n_X + 3n_P + 3n_\psi/2.$ %

This hierarchy is natural for the kinematic/dynamical and positivity constraints, as it satisfies the following properties:

\begin{enumerate}
    \item Acting with the supercharge increases the level by $1/2$: %
    \begin{equation}
        \ell( \{Q_\alpha, \mathcal{O}_\alpha\} ) = \ell(\mathcal{O}_\alpha) + \frac{1}{2}.
    \end{equation}
    This is expected since the supercharge $Q_\alpha$ is a level $\frac{7}{2}$ operator and the anti-commutator lowers the level by 3.
    
    \item For the kinematic constraints, the cyclicity condition in Eq.~\eqref{eq: cyc} is uniform in $\ell(\mathcal{O})$, except for the double-trace term from the commutator or anti-commutator, which has level $\ell(\mathcal{O}) - 3$. The Hermiticity/time-reversal condition relates operators of the same level, and the gauge condition \eqref{eq: gaugeConstraint} gives a level %
    $\ell(\mathcal{O}) + 3$ constraint.%

    \item To obtain the positivity conditions involving variables up to level $\ell_\text{cutoff}$, we need to select operators up to level $\frac{1}{2} \ell_\text{cutoff}$ and take their inner products.
\end{enumerate}

\begin{table}
\centering
\begin{tabular}{ |c|c|c| } 
\hline
 level cutoff   & total variables& free variables  \\ 
 \hline 
 4              & 11& 3                              \\ 
 \hline
 5              & 38& 4                                 \\
 \hline
 6              &140& 11                                \\
 \hline
 7              &569& 18                                \\
 \hline
 8              & 2528& 59                                \\
 \hline
 9              & 12077& 149                             \\
\hline
\end{tabular}
\caption{\label{tab: freevariables} The number of free variables in the semi-definite programming problem (after quotienting all the kinematic and dynamical constraints). We also report the number of variables before using the kinematic/dynamical constraints. For level 9, the computational time to solve the linear equations takes $\sim 1$ sec.}
\end{table}
To practically solve the bootstrap problem, we first symbolically solve the kinematic/dynamical constraints. That is, for a given level cutoff, we select a subset of ``free variables" and solve for all other variables below this level using this subset of free variables. In Table~\ref{tab: freevariables}, we show the number of free variables for each level truncation.\footnote{The practical choice of free variables is relatively arbitrary; below is the list of our choices up to level $6$ (for compactness, we omit the trace below):
\begin{small}
\begin{equation*}
\begin{split}
    &\left\langle P^{\text{I}}P^{\text{I}}\right\rangle ,\left\langle X^{\text{I}}X^{\text{I}}\right\rangle ,\gamma _{\beta \alpha }^{\text{I}\text{J}} \left\langle P^{\text{I}}X^{\text{J}}\psi _{\beta }\psi _{\alpha }\right\rangle ,\left\langle X^{\text{I}}X^{\text{I}}X^{\text{J}}X^{\text{J}}\right\rangle , \left\langle X^{\text{I}}X^{\text{I}}\psi _{\alpha }\psi _{\alpha }\right\rangle ,\left\langle \psi _{\beta }\psi _{\beta }\psi _{\alpha }\psi _{\alpha }\right\rangle ,\gamma _{\eta \epsilon }^{\text{I}}\gamma _{\beta \alpha }^{\text{I}} \left\langle \psi _{\eta }\psi _{\epsilon }\psi _{\beta }\psi _{\alpha }\right\rangle ,\gamma _{\beta \alpha }^{\text{J}} \left\langle X^{\text{I}}X^{\text{I}}X^{\text{J}}\psi _{\beta }\psi _{\alpha }\right\rangle ,\\
    &\gamma _{\beta \alpha }^{\text{J}} \left\langle X^{\text{I}}X^{\text{I}}\psi _{\beta }X^{\text{J}}\psi _{\alpha }\right\rangle ,\gamma _{\beta \alpha }^{\text{J}} \left\langle X^{\text{I}}X^{\text{J}}\psi _{\beta }X^{\text{I}}\psi _{\alpha }\right\rangle ,\left\langle X^{\text{I}}X^{\text{I}}X^{\text{J}}X^{\text{J}}X^{\text{K}}X^{\text{K}}\right\rangle .
    \end{split}
\end{equation*}
\end{small}
}
These solutions in terms of the ``free variables," together with the positivity condition, form a \textit{linear matrix inequality}, which is the most traditional version of semidefinite programming.

\subsection{Worked example: 4 fermions}\label{sec: example}
To illustrate the positivity + kinematic constraints more concretely, let us consider the correlation functions involving $\ev{\tr \psi_{\alpha_1} \psi_{\alpha_2} \psi_{\alpha_3} \psi_{\alpha_4}}$.
We start by decomposing the 2 fermions into irreps:
\begin{align}
{\bf 16} \otimes {\bf 16} &= {\bf 1} + {\bf 9} + {\bf 36} + {\bf 84} + {\bf 126} \label{decomp2f}\\
    \psi_\alpha \psi_\beta &=\frac{1}{2} \delta_{\alpha\beta} O+\frac{1}{16}\gamma^I_{\alpha\beta} O^I+\frac{1}{32}\gamma^{IJ}_{\alpha\beta} O^{IJ}+\frac{1}{96}\gamma^{IJK}_{\alpha\beta} O^{IJK}+\frac{1}{384}\gamma^{IJKL}_{\alpha\beta} O^{IJKL} \label{decomp2f2}
\end{align}
The irreps on the RHS of \nref{decomp2f} are the fully anti-symmetric tensors; the factors in \nref{decomp2f2} are purely conventional.
Thus the four fermion correlators are constrained to have the form
\begin{align}
\label{4fermion}
        \ev{\tr(\psi_\alpha \psi_\beta \psi_\eta \psi_\epsilon)}
    &={\red a_1} \delta_{\alpha \beta} \delta_{\eta \epsilon}+{\red a_9} \gamma_{\alpha \beta}^I \gamma_{\eta \epsilon}^I+ {\red a_{36}} \gamma_{\alpha \beta}^{I J} \gamma_{\eta \epsilon}^{I J}+ {\red a_{84}} \gamma_{\alpha \beta}^{I J K} \gamma_{\eta\epsilon}^{I J K}+{\red a_{126}} \gamma_{\alpha \beta}^{I J K L} \gamma_{\eta\epsilon}^{I J K L}
\end{align}
Now using cyclicity of the trace and the anti-commutation relations \nref{canonical}:
\begin{align}
   \ev{\tr \psi_\alpha \psi_\beta \psi_\eta \psi_\epsilon}  &=-\ev{\tr \psi_\beta \psi_\eta \psi_\epsilon \psi_\alpha} + \frac{1}{2}\delta_{\alpha\beta}\delta_{\eta\epsilon}+\frac{1}{2}\delta_{\alpha\epsilon}\delta_{\eta\beta}  .\label{4fermion_cyc}
\end{align}
Now we may also expand the first term in \nref{4fermion_cyc} terms of SO(9) blocks to obtain
\begin{align} \la{cyclicity4f}
    {\red a_1} \delta_{\alpha \beta} \delta_{\eta \epsilon}+{\red a_9} \gamma_{\alpha \beta}^I \gamma_{\eta \epsilon}^I+ {\red a_{36}} \gamma_{\alpha \beta}^{I J} \gamma_{\eta \epsilon}^{I J}+ {\red a_{84}} \gamma_{\alpha \beta}^{I J K} \gamma_{\eta\epsilon}^{I J K}+{\red a_{126}} \gamma_{\alpha \beta}^{I J K L} \gamma_{\eta\epsilon}^{I J K L} = \\
    \frac{1}{2} \delta_{\alpha\beta}\delta_{\eta\epsilon}+\frac{1}{2}\delta_{\alpha\epsilon}\delta_{\eta\beta}  - {\red a_1} \delta_{ \beta \eta} \delta_{\epsilon \alpha}-{\red a_9} \gamma_{ \beta \eta}^I \gamma_{\epsilon \alpha}^I- {\red a_{36}} \gamma_{ \beta \eta}^{I J} \gamma_{\epsilon \alpha}^{I J}-{\red a_{84}} \gamma_{ \beta \eta}^{I J K} \gamma_{\epsilon \alpha}^{I J K}- {\red a_{126}} \gamma_{ \beta \eta}^{I J K L} \gamma_{\epsilon \alpha}^{I J K L} 
\end{align}
Now to solve this equation, we need the crossing relations for the SO(9) blocks. We have introduced a graphical notation, where each vertex corresponds to a Clebsch-Gordan symbol, and internal lines represent sums over common indices. We want to expand an $s$-channel block in a basis of $t$-channel blocks:
\begin{align}
\begin{tikzpicture}[baseline={(current bounding box.center)},scale=0.9]
\draw[thick] (-1,1) node[left] {$\beta$} -- (0,0);
\draw[thick] (-1,-1) node[left] {$\alpha$} -- (0,0);
\draw[thick] (0,0) -- (1.5,0) node[midway, above] {$R_s$};
\draw[thick] (1.5,0) -- (2.5,1) node[right] {$\eta$};
\draw[thick] (1.5,0) -- (2.5,-1) node[right] {$\epsilon$};
\end{tikzpicture}
\quad = \quad \sum_{R_t} \; \mathbb{F}_{R_s,R_t}\left[\begin{array}{ll}
{\bf 16} & {\bf 16} \\
{\bf 16} & {\bf 16}
\end{array}\right]  \;
\begin{tikzpicture}[baseline={(current bounding box.center)},scale=0.9]
\draw[thick] (-1,1) node[left] {$\beta$} -- (0,0);
\draw[thick] (0,0) -- (1,1) node[right] {$\eta$};
\draw[thick] (0,0) -- (0,-1.25) node[midway, right] {$R_t$};
\draw[thick] (-1,-2.25) node[left] {$\alpha$} -- (0,-1.25);
\draw[thick] (0,-1.25) -- (1,-2.25) node[right] {$\epsilon$};
\end{tikzpicture}
\la{defCrossing}
\end{align}
Here we have defined the crossing kernel $\mathbb{F}_{R_s,R_t}$ for the case where the 4 operators are all in the spinor irreps. To compute the crossing kernel\footnote{We are using notation that will be familiar to conformal bootstrappers. There is a strong analogy: the conformal dimension, a label for the conformal group irrep, is analogous to the label of the irrep of SO(9). The conformal blocks are analogous to the SO(9) blocks. The coefficients $\red a$ are analogous to the sqaures of OPE coefficients $C^2$, or more precisely $C_{ijR} C_{Rkl}$ which satisfy analogous positivity requirements.}, one may simply note that it is an overlap between $s$ and $t$-channel blocks, e.g., it is a 6$j$-symbol for SO(9):
\begin{align}
    \mathbb{F}_{R_s, R_t}\left[\begin{array}{ll}
R_1 & R_2 \\
R_3 & R_4
\end{array}\right] &=  \frac{1}{\mathcal{N}} \; \scalebox{1.5}{$\Bigg \langle$} \begin{tikzpicture}[baseline={(current bounding box.center|-0,-0.4cm)},scale=0.65]
\draw[thick] (-1,0.5) node[left] {$R_1$} -- (0,0);
\draw[thick] (0,0) -- (1,0.5) node[right] {$R_2 $};
\draw[thick] (0,0) -- (0,-1) node[midway, right] {$R_t$};
\draw[thick] (-1,-1.5) node[left] {$ R_3 $} -- (0,-1);
\draw[thick] (0,-1) -- (1,-1.5) node[right] {$R_4$};
\end{tikzpicture} \scalebox{1.5}{$\Bigg |$} \begin{tikzpicture}[baseline={(current bounding box.center)},scale=0.6]
\draw[thick] (-1,1) node[left] {$R_1$} -- (0,0);
\draw[thick] (-1,-1) node[left] {$R_3$} -- (0,0);
\draw[thick] (0,0) -- (1.5,0) node[midway, above] {$R_s$};
\draw[thick] (1.5,0) -- (2.5,1) node[right] {$R_2$};
\draw[thick] (1.5,0) -- (2.5,-1) node[right] {$R_4$};
\end{tikzpicture}  \scalebox{1.5}{$\Bigg \rangle$} 
=  \frac{1}{\mathcal{N}} \;
\begin{tikzpicture}[scale=1.4, baseline={(current bounding box.center)}]
    \coordinate (A) at (0, 1);
    \coordinate (B) at (-0.9, 0);
    \coordinate (C) at (0.9, 0);
    \coordinate (D) at (0, -0.5);
    \draw[thick] (A) -- (B) node[midway, above left] {$R_1$};
    \draw[thick] (A) -- (C) node[midway, above right] {$R_2$};
    \draw[thick] (A) -- (D) node[midway, above right] {$R_t$};
    \draw[thick] (B) -- (D) node[midway, below left] {$R_3$};
    \draw[thick] (C) -- (D) node[midway, below right] {$R_4$};
    \draw[thick,dashed] (B) -- (C) node[midway, above left] {$R_s$};
\end{tikzpicture}
\end{align}
The tetrahedral graph instructs us on how to sum over the indices, with each vertex corresponding to a Clebsch-Gordan coefficient. Furthermore, we should divide by the norm of the state that appears on the RHS of \nref{defCrossing}:
\begin{align}
{\mathcal{N}} \; = \; \scalebox{1.5}{$\Bigg \langle$} \begin{tikzpicture}[baseline={(current bounding box.center|-0,-0.4cm)},scale=0.65]
\draw[thick] (-1,0.5) node[left] {$R_1$} -- (0,0);
\draw[thick] (0,0) -- (1,0.5) node[right] {$R_2 $};
\draw[thick] (0,0) -- (0,-1);
\draw[thick] (-1,-1.5) node[left] {$ R_3 $} -- (0,-1);
\draw[thick] (0,-1) -- (1,-1.5) node[right] {$R_4$};
\node at (0.5,-0.5) {$R'$};
\end{tikzpicture} \scalebox{1.5}{$\Bigg |$} \begin{tikzpicture}[baseline={(current bounding box.center|-0,-0.4cm)},scale=0.65]
\draw[thick] (-1,0.5) node[left] {$R_1$} -- (0,0);
\draw[thick] (0,0) -- (1,0.5) node[right] {$R_2 $};
\draw[thick] (0,0) -- (0,-1);
\draw[thick] (-1,-1.5) node[left] {$ R_3 $} -- (0,-1);
\draw[thick] (0,-1) -- (1,-1.5) node[right] {$R_4$};
\node at (0.5,-0.5) {$R'$};
\end{tikzpicture} \scalebox{1.5}{$\Bigg \rangle$}.
\end{align}
For the case where all the external irreps are spinors, the above formulas give: %
\begin{align}
  \mathbb{F}_{R_s, R_t}\left[\begin{array}{ll}
{\bf 16} & {\bf 16} \\
{\bf 16} & {\bf 16}
\end{array}\right] = \left(
\begin{array}{ccccc}
 \frac{1}{16} & \frac{1}{16} & -\frac{1}{32} & -\frac{1}{96} & \frac{1}{384} \\
 \frac{9}{16} & -\frac{7}{16} & -\frac{5}{32} & \frac{1}{32} & \frac{1}{384} \\
 -\frac{9}{2} & -\frac{5}{2} & \frac{1}{2} & 0 & \frac{1}{48} \\
 -\frac{63}{2} & \frac{21}{2} & 0 & \frac{1}{2} & \frac{1}{16} \\
 189 & 21 & \frac{21}{2} & \frac{3}{2} & \frac{3}{8} \\
\end{array}
\right)
\end{align}
Here each row/column corresponds to an irrep, organized by increasing irrep dimension.
One can check that this matrix satisfies the property $\mathbb{F}^2 = 1$.
For the case of spinors, these identities are usually referred to as Fierz identities. We have given a somewhat long-winded explanation of how to reproduce these identities, since the full machinery will be necessary when we consider higher irreps.

With this crossing kernel, we can then solve \nref{cyclicity4f}. This means we may eliminate 3 of the 5 variables. However, that positivity of these three variables then leads to constraints on the other two:
\begin{align} \label{positivity4f1}
\frac{9!}{5!} {\red a_{126}} &= \frac{1}{6144} \tr (O^{IJKL} O^{IJKL}) = \frac{1}{672} \left(-2 {\red a_1}-2 {\red a_9}+1\right) \ge 0,\\ \label{positivity4f2}
{\frac{9!}{6!}\red a_{84}} &=  \frac{1}{1536} \tr (O^{IJK} O^{IJK}) = \frac{10 {\red a_1}-18 {\red a_9}-5}{1008} \le 0,\\ \label{positivity4f3}
{\frac{9!}{7!}\red a_{36}} &= \frac{1}{512}\tr (O^{IJ} O^{IJ}) = \frac{1}{48} \left(2 {\red a_1}+6 {\red a_9}-1\right) \le 0,\\ \label{positivity4f4}
& 9 {\red a_9} = \frac{1}{256}  \ev{\tr O^I O^I} \ge 0, \quad \left(
\begin{array}{cc}
 1 & 8 \\
 8 & 256 {\red a_1} \\
\end{array}
\right) \succeq 0  
\end{align}
In the matrix inequality \nref{positivity4f4}, we used the fact that the identity is also an SO(9) singlet, and that $\ev{\tr \psi_\alpha \psi_\alpha} = 8$.
Optimizing for $a_1$, we get 
\begin{align} \label{level6bd}
    1 \le \ev{\tr O O} \le 2
\end{align}

Actually, by repeating this exercise for 6 fermions (level 9), we can improve the bound to $1 \le \ev{\tr OO} \le 1.53125 $. To go to higher levels, we need the crossing relations not just for the 4-pt SO(9) blocks but the higher-pt blocks.

\color{black}

\section{Bootstrap results}

\begin{figure}[t]
\centering
\includegraphics[width=0.850\textwidth]{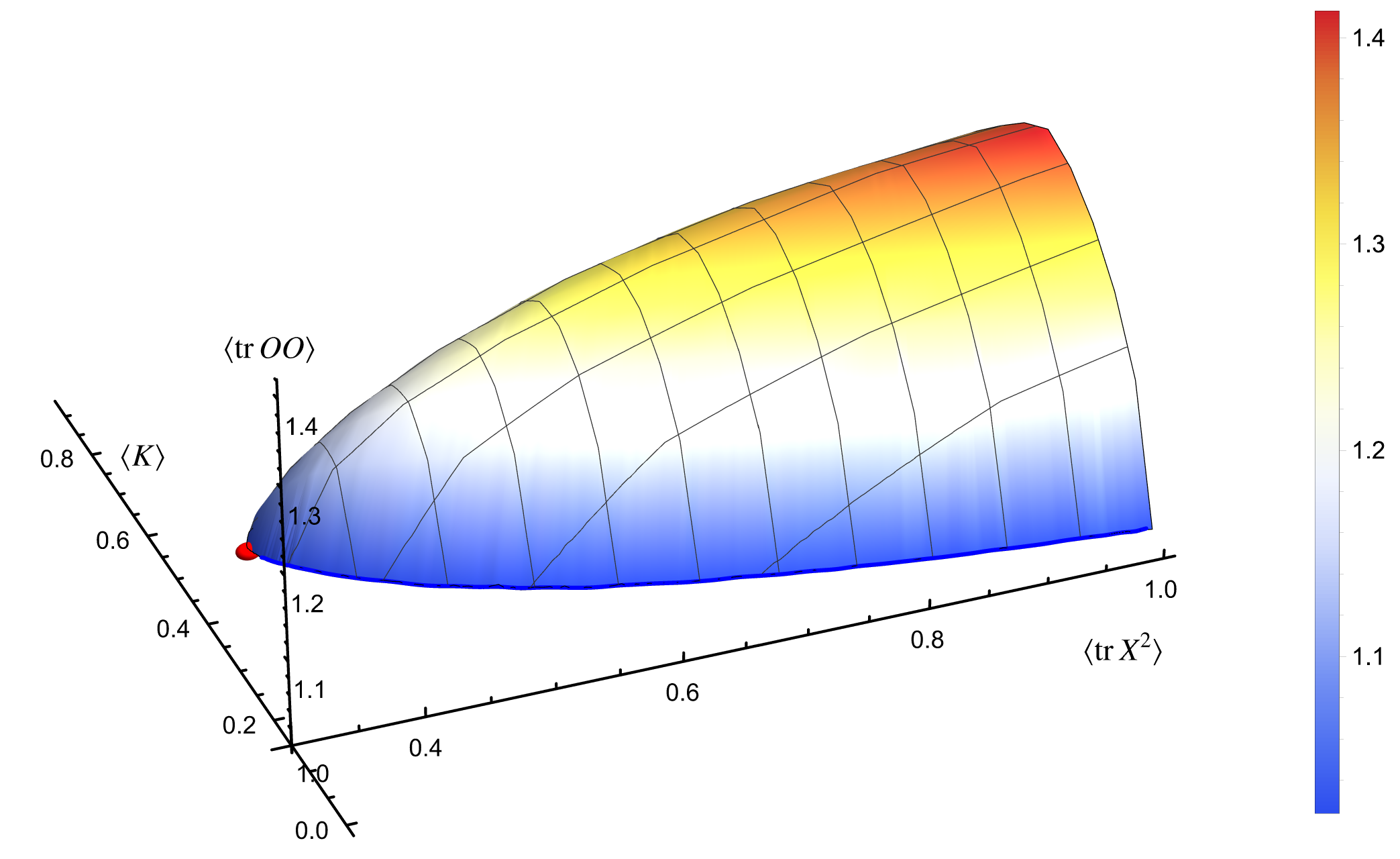}
\caption{\label{fig: plot3d}{ This 3D figure shows the shoe-like allowed region for the level $9$ bootstrap. The lower bound of $\langle \tr O O\rangle$ is not visible from the plot, but it is actually very flat, uniformly slightly above $1$, which means slightly better than the purely kinematic constraint, equation \eqref{level6bd}, of $\langle \tr O O\rangle$. The tip of the shoe (the {\red red dot}) represents one of our estimates of the variables from the bootstrap: $(\langle \tr X^2\rangle,\langle K\rangle, \langle \tr O O\rangle)\approx(0.355,0.495,1.02)$.}  }
\end{figure}
Using the method described in the previous section, we perform the bootstrap procedure up to level $9$. To present the main result, we introduce the short-hand notation:
\begin{equation}
    \langle \tr O O\rangle=\frac{1}{64}\langle \tr \psi_\alpha\psi_\alpha\psi_\beta\psi_\beta\rangle,\quad \langle \tr X^2\rangle=\frac{1}{9}\langle \tr X^I X^I\rangle,\quad \langle \tr X^2 X^2\rangle=\frac{1}{81}\langle \tr X^I X^I X^J X^J\rangle
\end{equation}
and the normalized kinetic/potential term
\begin{equation}
    \langle K\rangle=\frac{1}{18}\langle \tr P^I P^I\rangle=-\frac{1}{36}\langle \tr [X^I, X^J]^2\rangle = -\frac{1}{18} \tr  (\psi_\alpha \psi_\beta  X^I) \gamma^I_{\alpha \beta} 
\end{equation}
Up to level 9, the relevant fermionic irreps are $\{ {\bf 16}, {\bf 128}, {\bf 432}, {\bf 576}, {\bf 672}, {\bf 768}, {\bf 1920}, {\bf 2560} \}$ in addition to bosonic irreps corresponding to all Young tableaux generated corresponding to partitions of integers $\{1,2,3,4\}$ and the fully anti-symmetric tensor of rank 5. %
We also note that at this level, we have a standard SDP problem if we scan over the value of $\ev{\tr X^2}$, e.g., after using large $N$ factorization, for fixed $\ev{\tr X^2}$ the equations are linear in all other variables. At higher levels, the constraints will become non-linear in more variables, for example at level 11, we expect level 4 operators such as $K$ to enter quadratically due to the cyclicity constraints. For higher levels, one can either scan over more variables or use the non-linear relaxation method \cite{Kazakov:2021lel}. The latter is likely to be preferable beyond level 11. 

\begin{figure}[t]
\centering
\includegraphics[width=0.950\textwidth]{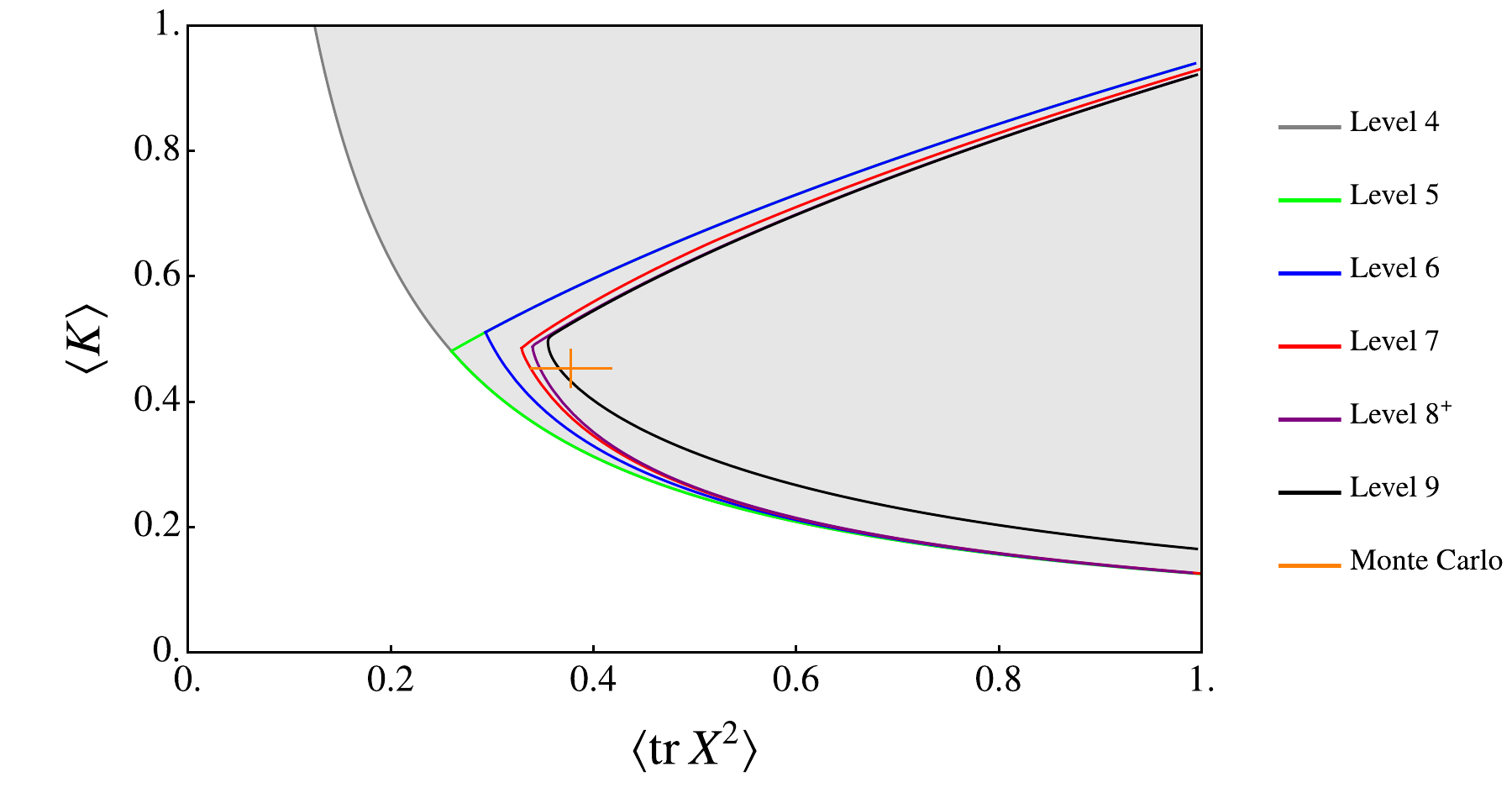}

\caption{\label{fig: plconvergence}{The allowed region is projected onto the $\langle \tr X^2 \rangle$-$\langle K \rangle$ plane, with bootstrap levels ranging from $4$ to $9$. At level $4$, there is no upper bound for $\langle K \rangle \propto \ev{\tr P^2} \propto \ev{\tr [X^I, X^J]^2}$, while for higher levels, for each given value of $\langle \tr X^2 \rangle$, we obtain both an upper and a lower bound for $\langle K \rangle$. When the given upper and lower bounds meet, we establish a lower bound for $\langle \tr X^2 \rangle$ at the corresponding level. We also show the extrapolation of the Monte Carlo \cite{Berkowitz:2018qhn} results, see Appendix \ref{extrapolate} for details. }
}
\end{figure}

Figure~\ref{fig: plot3d} shows the allowed region for the level $9$ bootstrap. Up to this level, the allowed region is not compact, meaning that we do not obtain an unconditional upper bound on $\langle \tr X^2 \rangle$. However, we do have a uniform bound for $\langle \tr O O \rangle$, and for a given value of $\langle \tr X^2 \rangle$, the expectation value of the kinematic energy is bounded in both directions. There is strong evidence that the lower bound on $\langle \tr X^2 \rangle$ (and more generally, the tip of the peninsula) is converging to a limit point. This is illustrated in Table~\ref{tab: lowerbound} and Figure~\ref{fig: plconvergence}. In the future, it would be interesting to compare with Monte Carlo results for $\ev{\tr OO}$ or $\ev{\tr O^I O^I}$. A measurement of these fermionic correlators that is consistent with the bootstrap bounds would be a highly non-trivial test of the conjecture that the sign problem is unimportant even at low temperatures \cite{Catterall:2008yz, Catterall:2009xn, Berkowitz:2016jlq}. Alternatively, by assuming there is no sign problem at low temperatures, the Monte Carlo result can be used to test whether the true answer lies close to the tip of the peninsula.

\begin{table}[t]
\centering
\begin{tabular}{ |c|c| } 
\hline
 method & $\langle \tr X^2 \rangle $   \\ 
 \hline 
 \makecell{Monte Carlo \cite{Berkowitz:2018qhn}   } & $\approx 0.378 \pm 0.04$  \\ 
 \hline
  \makecell{Monte Carlo \cite{Pateloudis:2022ijr} } & $[0.346, 0.430]$ \\
 $T = 0.35, \mu =0.5$ & $0.399 \pm 0.001 $
 \\
 $T = 0.4, \mu =0.5$ & 
 $0.394 \pm 0.001$
 \\
 \hline
 \makecell{primitive bootstrap\\ \cite{Lin:2023owt} }   & $\ge 0.1875$   \\
  \hline
   \makecell{bootstrap\\ level 5}     & $\ge 0.260$   \\
  \hline
 \makecell{bootstrap\\ level 6}     & $\ge 0.294$   \\
  \hline
 \makecell{bootstrap\\ level 7}     & $\ge 0.329$   \\ 
  \hline
\makecell{bootstrap\\ level 8$^+$}     & $\ge 0.340$   \\ 
\hline
 \makecell{bootstrap\\ level 9}     & $\ge 0.355$ \\
 \hline
\end{tabular}
\caption{\label{tab: lowerbound}{lower bounds on $\ev{\tr X^2}$ at different levels of the bootstrap, compared with the Monte Carlo results of \cite{Berkowitz:2018qhn, Pateloudis:2022ijr}. For \cite{Berkowitz:2018qhn} we performed an extrapolation to estimate the error bars, see Appendix \ref{extrapolate} for details. For \cite{Pateloudis:2022ijr}, we report the lowest temperature $T=0.3, N = 16, L=24$ result, with the upper value in the range corresponding to the $\mu=0.3$ result and the lower value corresponding to the $\mu=0.8$ result. We also report the large $N$ and continuum extrapolation performed by Stratos Pateloudis \cite{Pateloudis:2022ijr} for the $T=0.35$ and $T=0.4$ data at $\mu = 0.5$.  } }
\end{table}

In Table~\ref{tab: lowerbound}, we show the lower bound at each bootstrap level and compare it with the Monte Carlo results~\cite{Berkowitz:2016jlq, Pateloudis:2022ijr} (with extrapolation)\footnote{The Monte Carlo value of $\ev{\tr X^2}$ {\it decreases} slightly as temperature increases. This seems to indicate the difficulty of isolating the bound state from the continuum at finite $N$.}. Figure~\ref{fig: plconvergence} shows the allowed region (the peninsula enclosed by different colors) of $(\langle \tr X^2 \rangle, \langle K \rangle)$ from level $4$ to level $9$. The lower bounds listed in Table~\ref{tab: lowerbound} correspond to the peninsula's tip for each color. We note that level $8^+$ here refers to all the level $8$ constraints plus the six $\psi$ correlator constraints (similar to the four $\psi$ situation discussed in Section~\ref{sec: example}), which are part of the level $9$ constraints. The reason for not considering the pure level $8$ is that it seems the pure level $8$ constraints do not provide any improvement over the level $7$ bound on $\langle \tr X^2 \rangle$.

Furthermore, one can derive exact analytic bootstrap bounds at low levels due to the simplicity of the corresponding optimization problem. For example, the upper and lower bounds for the level $6$ bootstrap (displayed in Figure \ref{fig: plconvergence}) are given by:
\begin{align}
    \langle \tr [X^I, X^J]^2 \rangle &\le \frac{4}{3} \sqrt{2 \langle \tr X^2 \rangle }, \\
    3 \langle \tr [X^I, X^J]^2 \rangle^3 &- 64 \langle \tr [X^I, X^J]^2 \rangle \langle \tr X^2 \rangle + 16 \leq 0.
\end{align}
The lower bound of $\langle \tr X^2 \rangle$ (see Table \ref{tab: lowerbound}) comes from the crossing of these two curves:
\begin{align}
    \langle \tr X^2 \rangle \geq \frac{3}{4} \left( \frac{3}{50} \right)^{1/3} \approx 0.2936.
\end{align}

\begin{figure}[t]
\centering
\includegraphics[width=0.950\textwidth]{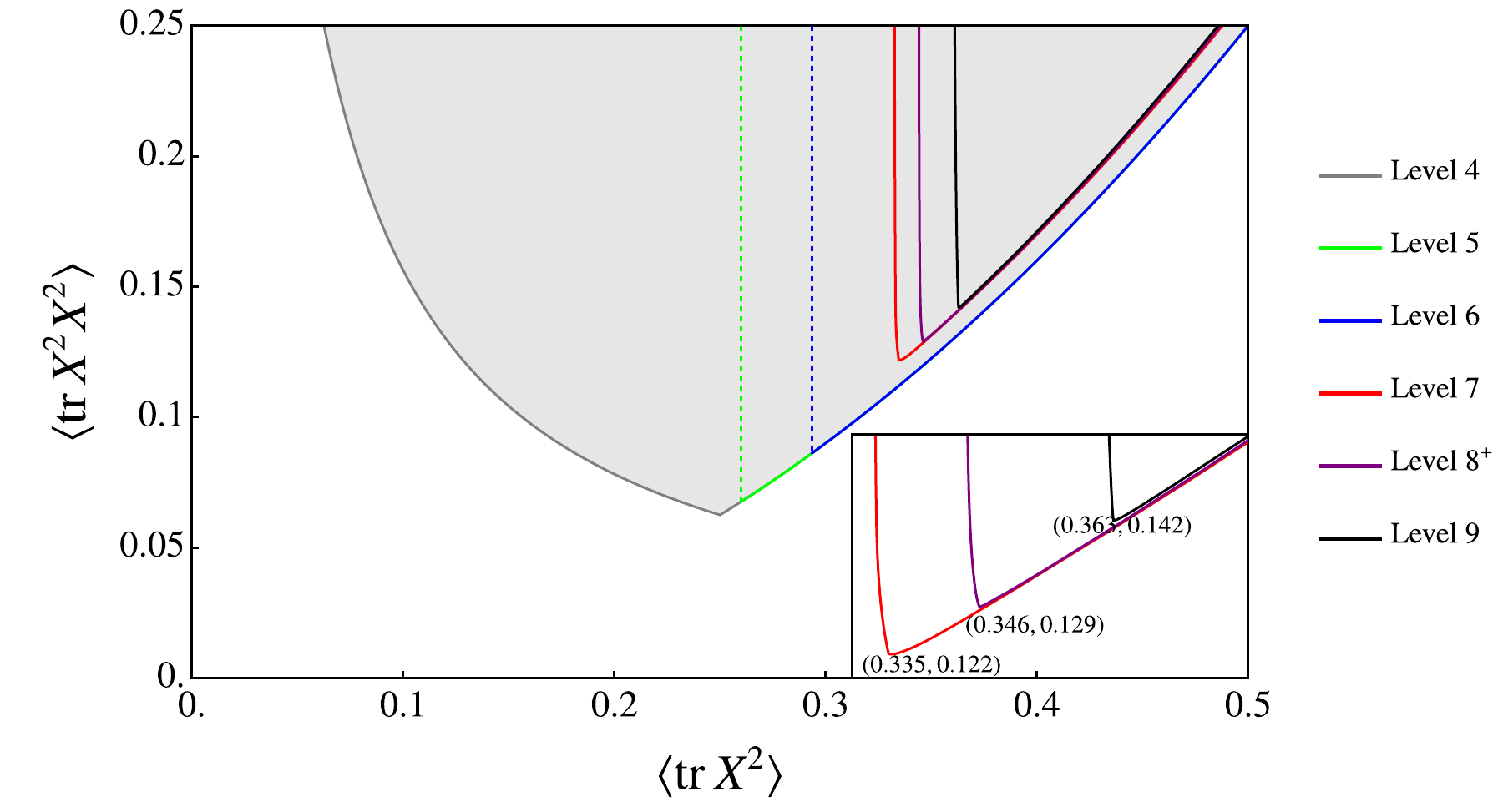}
\caption{\label{fig: plx4}{Bound for $\langle \tr X^2 X^2 \rangle$: for the {\color{green} level 5} and {\color{blue} level 6} bounds, there is a critical value of $\langle\tr X^2\rangle$ (corresponding to the tip of the peninsula in Figure \ref{fig: plconvergence} where the allowed region ends). This is not the case for the higher level constraints, where the minimal allowed value of $\langle\tr X^2\rangle$ can only be achieved by smoothly sending $\langle\tr X^2 X^2\rangle$ to $+\infty$. This behavior is illustrated in the inset zoomed-in plot, where a very steep (but not vertical!) lower bound for $\langle\tr X^2 X^2\rangle$ appears as we approach the tip of the peninsula in Figure~\ref{fig: plconvergence}}. The level 4 constraint was essentially derived by Polchinski \cite{Polchinski:1999br}.}
\end{figure}

The reader may wonder what the allowed region for longer-length operators.
We consider this in Figure~\ref{fig: plx4}. In this figure, we plot the lower bound of $\langle \tr (X^2)^2 \rangle$ for a given $\langle \tr X^2 \rangle$. Of course, since there is no unconditional upper bound on $\ev{\tr X^2}$ up to level 9, it follows that there is also no upper bound on $\langle \tr (X^2)^2 \rangle$.

Perhaps more interesting is that starting from level $7$, a kink appears when the value of $\langle \tr X^2 \rangle$ is close to the strict lower bound at the corresponding level (the tip of the peninsula in Figure \ref{fig: plconvergence}). The explanation is straightforward: to approach the lower bound on $\ev{\tr X^2}$, an extremely large value of $\langle \tr X^2 X^2 \rangle$ is needed. %

In fact, the variables $\langle \tr (X^2)^m \rangle$ with $m\geq 2$ exhibit interesting behavior in our current bootstrap setup. They completely decouple from the dynamical constraints in equation ~\eqref{groundBoot}, and when we approach the strict lower bound of $\langle \tr X^2 \rangle$ for the corresponding level, the operators $\langle \tr (X^2)^m \rangle$ tend to diverge. This seems to be closely related to the existence of flat directions in the potential energy $\propto [X^I, X^J]^2$ in Eq.~\eqref{ham}. %
These operators may lead to serious numerical instabilities when probing the lower bound of $\langle \tr X^2 \rangle$. We developed a technique to identify and remove unbounded variables associated with these flat directions, which we will detail in \cite{LinZheng2}. %

Finally, we want to emphasize that all the numerical results presented so far only required very modest computational resources. A 2021 M1-Max MacBook Pro was sufficient to generate all the figures in the current section from scratch within one hour. Most of the computational time was spent performing the gamma matrix algebra (which could be further optimized), rather than solving the optimization problem. This stands in sharp contrast to recent Monte Carlo simulations \cite{Pateloudis:2022ijr}, which took weeks of time on a cluster. Of course, we have only derived {\it bounds} on correlation functions as opposed to their value; in the absence of bootstrap ``islands,'' a comparison of the efficiency of the bootstrap method and Monte Carlo is not strictly speaking possible unless one grants the additional assumption that the true physical theory lies close to a ``kink''. With this assumption, one can then assert that the bootstrap method is more efficient at producing ground state correlators at infinite $N$. Of course, it is thanks to Monte Carlo  that we have evidence for this assumption! %

\section{Discussion \label{sec: discuss}}

In this section, we clarify some subtleties associated with the large-$N$ limit and give some future directions.

We have imposed infinite $N$ into the bootstrap constraints by using large-$N$ factorization in \nref{eq: cyc}. This is equivalent to assuming that off-diagonal elements are suppressed, e.g., 
\begin{align}\la{factorization}
\bra{\Omega}\tr \O_1 \tr \O_2\ket{\Omega} = \int \diff{E} \, \rho(E) \bra{\Omega}\tr \O_1 \ketbra{E} \tr \O_2\ket{\Omega}  \approx  \bra{\Omega}\tr \O_1 \ketbra{\Omega} \tr \O_2\ket{\Omega}. 
\end{align}
Justifying this equation is somewhat subtle, and indeed we believe this approximation can fail for sufficiently complicated operators $\O$. 
The usual argument for factorization relies on the 't Hooft limit, where planar diagrams dominate. However, recall that in $d <4$ spacetime dimensions, the super Yang-Mills interaction is relevant. Hence the 't Hooft expansion is really controlled by $g^2 N/E^3$ where $E$ is a characteristic energy scale. Since we are studying the ground state $E=0$, the dimensionless 't Hooft parameter is diverging. However, what is important for the approximation \nref{factorization} is that the matrix elements with intermediate energies $E$ in the integral are really negligible. 
These intermediate energies are associated with finite 't Hooft coupling, which justifies the factorization.

The loophole is if the intermediate energies are very low (suppressed by powers of $N$). We expect that this exception is relevant for operators\footnote{Our notation is a bit sloppy, since for general integer $\ell$ there are many possible SO(9) invariant tensors. It would be nice to characterize the subset of these tensors lead to divergent operators.} such as  $\ev{\tr X^\ell}$. For $\ell < 9 $, we expect that the moments $\ev{\tr X^\ell}$ are finite  \cite{Polchinski:1999br} but due to the expected power law tail\footnote{See equation 7.10 in \cite{Polchinski:1999br} and also \cite{Plefka:1997xq, Frohlich:1999zf, Hoppe:2000tj,  Hasler:2002wt, Lin:2014wka}. } in the wavefunction associated with the flat directions, $\ev{\tr X^\ell \tr X^\ell }$ should diverge for $2 \ell \ge 9$. Thus, strictly speaking, large $N$ factorization is violated for these operators. Actually, if we consider other multi-traces, large $N$ factorization should fail even for $\mathcal{O} = \tr X^2$, since correlators like $\ev{(\tr X^2)^5}$ or $\ev{\tr X^2 \tr X^8}$ should also diverge.

We expect that all these divergences are $\sim 1/N^{p}$ effects, e.g., that the normalization of the power law tail in the wavefunction is suppressed at large $N$. 
To be a bit more precise, we can imagine adding a small BMN mass term. We expect that all the perturbative in $1/N^2$ corrections to any multi trace correlator would then be finite. However, in the $\mu \to 0 $ limit some $1/N^{p}$ corrections could have a diverging coefficient. If we take $N \to \infty$ first, and then $\mu \to 0$, we would extract the finite piece. This is presumably the relevant order of limits that one obtains by imposing large $N$ factorization and working directly with the $\mu = 0$ theory.

If the asymptotic wavefunction is suppressed by the appropriate powers of $1/N$, it is conceivable that the infinite $N$ bootstrap will yield finite estimates for the correlators $\ev{\tr X^\ell}$. In this scenario, large $N$ factorization would automatically pick out the finite, leading in $N$ contribution to the correlator $\ev{\tr X^\ell}$ even for $\ell > 9$. But the finite $N$ bootstrap would yield divergent correlators for $\ell >9$. This means that as we increase the level of the bootstrap, any finite value for $\tr X^\ell$ will be eventually ruled out for $\ell >9$. Whether or not this will actually happen is an interesting open question. Recently, the bootstrap constraints have been explored at finite $N$ in Yang-Mills theory, see \cite{Kazakov:2024ool}; presumably enforcing finite $N$ trace relations would allow us to also bootstrap the D0-brane quantum mechanics at finite $N$.

Another obvious question related to the presence of flat directions is whether the bootstrap will yield an upper bound on $\ev{\tr X^2}$ (and other low level correlators). At a minimum, we expect that by bootstrapping the BMN model \cite{Berenstein:2002jq} at small $\mu$ and infinite $N$ that we will eventually recover an upper bound. We conjecture that at any finite value of $\mu > 0$, the bootstrap should eventually converge, e.g., all islands in that model should shrink to a point. This makes it plausible that we should also get islands for $\mu = 0$. We also bootstrapped a model which has previously been discussed as a toy model for D0 branes (although it does not have flat directions at the quantum level), see Appendix \ref{app: toy}. At low levels, the peninsula in Figure \ref{fig: toyground} vaguely resembles the peninsula of Figure \ref{fig: plconvergence}, but at higher levels the peninsula becomes an island which is relatively close to the tip of the peninsula.

In general, operators which are finite in the
't Hooft limit are expected to have a 10D supergravity interpretation \cite{Itzhaki:1998dd}. %
For example, \cite{Polchinski:1999br, Lin:2023owt} observed that the typical size of the matrices in the ground state is parametrically of the same order as the size of the supergravity region (see also \cite{Hanada:2021ipb}). It would be interesting to study these observables at finite microcanonical energies \cite{Lin:2023owt} and/or finite temperature, building on the work of \cite{Fawzi:2023fpg, Cho:2024kxn}. In Appendix \ref{app: gpositivity}, we show that the zero temperature bootstrap condition in \cite{Fawzi:2023fpg} is actually redundant when we impose the supercharge condition.

On the other hand, from the 11D point of view, one can view these observables as constraints on the ground state wavefunction of a supergraviton. An upper bound on $\ev{\tr X^2}$ would be a concrete demonstration of a normalizable ground state. It would be interesting to explore how and if constraints on the ground state correlators could constrain the S-matrix of BFSS.

Finally, recall that the D0-brane quantum mechanics can be viewed as the dimensional reduction of 10 dimensional $\mathcal{N}=1$ super Yang-Mills (SYM) to 1D, or 4D $\mathcal{N}=4$ SYM to 1D. It is thus a simple example of a Yang-Mills theory with (adjoint) fermions. 
Our results show that the bootstrap method can be effective even when there is potential a sign problem in the Euclidean theory. It would be interesting to explore bootstrapping other gauge theories (with fermions and/or turning on a $\theta$ angle) in the Hamiltonian approach, building on the Euclidean approach of \cite{Anderson:2016rcw, Kazakov:2022xuh, Kazakov:2024ool}.

\section*{Acknowledgments}

We thank Minjae Cho, Barak Gabai, Masanori Hanada, Shota Komatsu, Juan Maldacena, João Penedones, Joshua Sandor, Stephen Shenker, Douglas Stanford, Xi Yin, and Xiang Zhao for the discussions. We thank Ulf Gran for sharing with us the code for \texttt{GAMMA}. 
H.L. is supported by a Bloch Fellowship and by NSF Grant PHY-2310429. Z.Z. is supported by Simons Foundation grant \#994308 for the Simons Collaboration on Confinement and
QCD Strings.
This research was supported in part by grant NSF PHY-2309135 to the Kavli Institute for Theoretical Physics (KITP). We also thank CERN for its hospitality where some of this research was carried out.

\appendix
\def\i{\mathrm{i}}

\section{Redundancy of ground state positivity for supersymmetric systems \label{app: gpositivity}}

A well-known positivity condition for the ground state reads:
\begin{equation}\label{eq: groundpos}
    \langle \mathcal{O}^\dagger[H, \mathcal{O}]\rangle\geq0,\quad \forall \mathcal{O}
\end{equation}
which encodes the fact that the expectation value of the Hamiltonian under the state $|\Omega \rangle$ is always lower than that under $\mathcal{O} |\Omega \rangle$ for any $\mathcal{O}$.
In this appendix, we show that this positivity condition is redundant for supersymmetric quantum systems once we impose the inner product positivity $\langle \mathcal{O}^\dagger \mathcal{O} \rangle$ for both fermionic and bosonic operators.

For our current system, we have\footnote{The factor $8$ may vary depending on the number of supercharges. The argument works for any supersymmetric quantum mechanics with at least one supercharge.}:
\begin{equation}
    Q_\alpha Q_\alpha= 8H 
\end{equation}
To show the redundancy of the ground state positivity condition, we will consider the cases where $\mathcal{O}$ is bosonic or fermionic separately. If $\mathcal{O}$ is an arbitrary bosonic operator, we have imposed by the supercharge equation:
\begin{equation}
    \begin{split}
        0=\ev{\{Q_\alpha, \mathcal{O}_{ji}^\dagger [Q_\alpha, \mathcal{O}_{ij}]\}}
    \end{split}
\end{equation}
Expanding the right-hand-side,
\begin{equation}
    \begin{split}
        0&=\ev{\{Q_\alpha, \mathcal{O}_{ji}^\dagger [Q_\alpha, \mathcal{O}_{ij}]\}}\\
&=\ev{([Q_\alpha, \mathcal{O}^\dagger])_{ik}([Q_\alpha, \mathcal{O}])_{ki}+\mathcal{O}_{ji}^\dagger \{Q_{\alpha},[Q_\alpha, \mathcal{O}_{ij}]\}}\\
&=\ev{([Q_\alpha, \mathcal{O}^\dagger])_{ik}([Q_\alpha, \mathcal{O}])_{ki}+\mathcal{O}_{ji}^\dagger\big( Q_\alpha(Q_\alpha \mathcal{O}_{ij}- \mathcal{O}_{ij} Q_\alpha)+(Q_\alpha \mathcal{O}_{ij}- \mathcal{O}_{ij} Q_\alpha) Q_\alpha \big)}\\
&=\ev{([Q_\alpha, \mathcal{O}^\dagger])_{ik}([Q_\alpha, \mathcal{O}])_{ki}+8 \mathcal{O}_{ji}^\dagger [H, \mathcal{O}_{ij}]}
    \end{split}
\end{equation}

Thus we have:
\begin{equation}
    \ev{\mathcal{O}_{ji}^\dagger [H, \mathcal{O}_{ij}]}=\frac{1}{8}\ev{([Q_\alpha, \mathcal{O}])_{ki}^\dagger([Q_\alpha, \mathcal{O}])_{ki}}\geq 0
\end{equation}

Similarly, for an arbitrary fermionic operator $\mathcal{O}$, we have imposed:
\begin{equation}
    \begin{split}
        0=\ev{\{Q_\alpha, \mathcal{O}_{ji}^\dagger \{Q_\alpha, \mathcal{O}_{ij}\}\}}
    \end{split}
\end{equation}
Expanding the right-hand-side,
\begin{equation}
    \begin{split}
        0&=\ev{\{Q_\alpha, \mathcal{O}_{ji}^\dagger \{Q_\alpha, \mathcal{O}_{ij}\}\}}\\
&=\ev{(\{Q_\alpha, \mathcal{O}^\dagger\})_{ik}(\{Q_\alpha, \mathcal{O}\})_{ki}-\mathcal{O}_{ji}^\dagger [Q_{\alpha},\{Q_\alpha, \mathcal{O}_{ij}\}]}\\
&=\ev{(\{Q_\alpha, \mathcal{O}^\dagger\})_{ik}(\{Q_\alpha, \mathcal{O}\})_{ki}-\mathcal{O}_{ji}^\dagger\big( Q_\alpha(Q_\alpha \mathcal{O}_{ij}+ \mathcal{O}_{ij} Q_\alpha)-(Q_\alpha \mathcal{O}_{ij}+ \mathcal{O}_{ij} Q_\alpha) Q_\alpha \big)}\\
&=\ev{(\{Q_\alpha, \mathcal{O}^\dagger\})_{ik}(\{Q_\alpha, \mathcal{O}\})_{ki}-8 \mathcal{O}_{ji}^\dagger [H, \mathcal{O}_{ij}]}
    \end{split}
\end{equation}

So we have 
\begin{equation}
    \ev{\mathcal{O}_{ji}^\dagger [H, \mathcal{O}_{ij}]}=\frac{1}{8}\ev{(\{Q_\alpha, \mathcal{O}\})_{ki}^\dagger(\{Q_\alpha, \mathcal{O}]\})_{ki}}\geq 0
\end{equation}

In conclusion, the ground state positivity condition, Eq.~\eqref{eq: groundpos}, is equivalent to the inner-product positivity of the $Q_\alpha$-exact operators.

\section{Bootstrapping a toy model \label{app: toy}}
In this appendix, we consider bootstrapping the ``toy supermembrane" Hamiltonian~\cite{deWit:1988xki, Frohlich:1999zf, Balthazar:2016utu, Komatsu:2024vnb}:
\begin{align}\label{eq: toyH}
    H = \frac{1}{2}(p_x^2 + p_y^2) + \frac{1}{2} g x^2 y^2
\end{align}
Here $p_x$, $p_y$ are the canonical conjugate of $x$ and $y$, respectively. This toy model shares many interesting features with the BFSS model, Eq.~\eqref{ham}, considered in the main text. For both models, we can set $g = 1$ by rescaling the variables without loss of generality. Additionally, both models lack a quadratic mass term, which poses challenges for perturbation theory and Hamiltonian truncation. We also note that a similar hierarchy can be established as in Section~\ref{sec: hierarchy}, by defining a level 
\begin{equation}
    \begin{split}
        \ell(x)=\ell(y)=1,\quad \ell(p_x)=\ell(p_y)=2.
    \end{split}
\end{equation}

\subsection{Arbitrary eigenstates}\label{sec: oldschool}
\begin{figure}[t]
\centering
\includegraphics[width=0.70\textwidth]{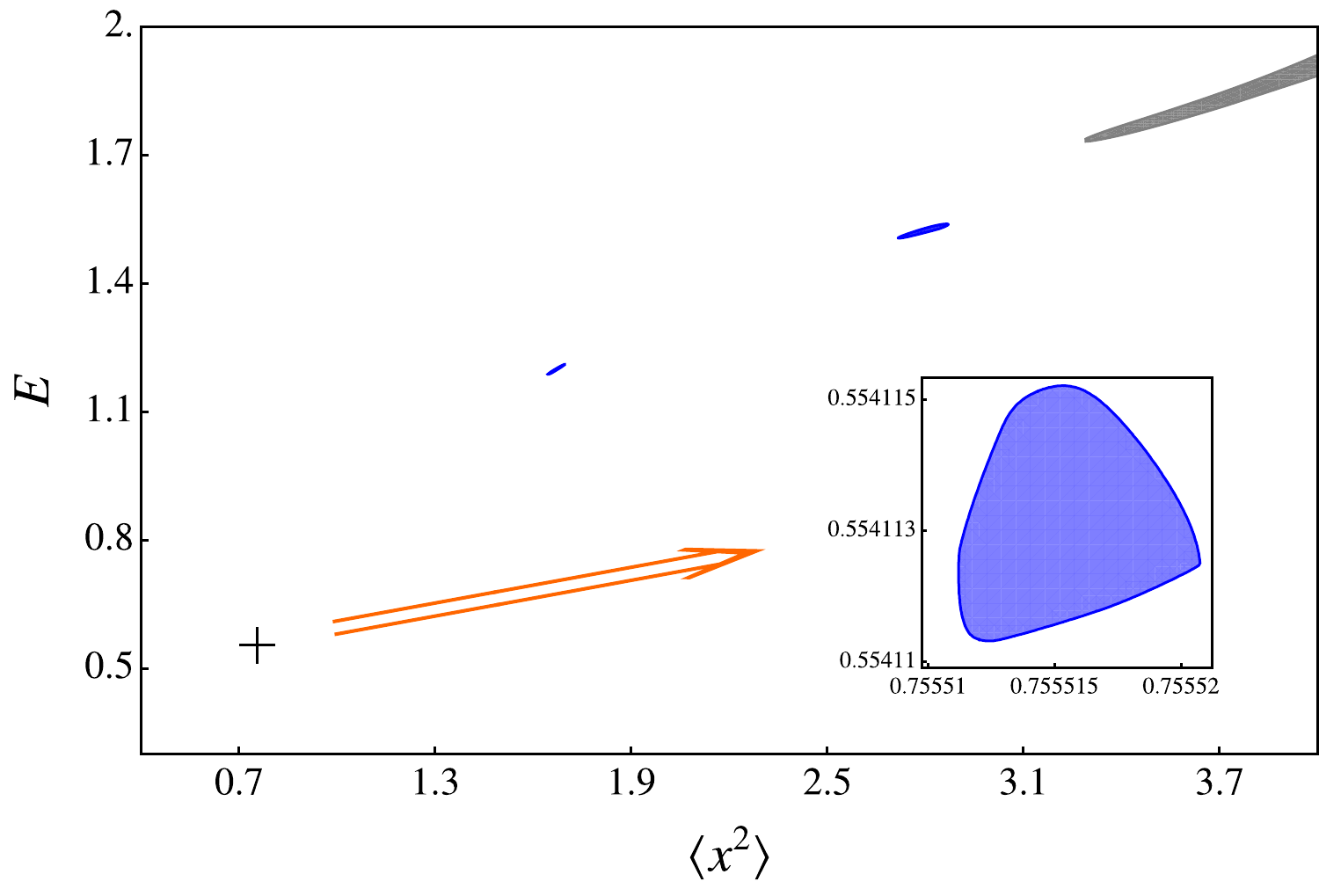}
\caption{\label{fig: x2y2}{}
Bootstrap ``archipelago'' bounds on the energy and $\ev{x^2}$ up to level $12$. We see that multiple energy eigenvalues above the ground state are resolved (before eventually being lost in the peninsula). }
\end{figure}

Suppose we have an eigenstate with energy $E$, and we impose the following condition on an arbitrary operator $\mathcal{O}$:
\begin{equation}
    \begin{split}
        \langle \mathcal{O} H \rangle = \langle H \mathcal{O} \rangle = E \langle \mathcal{O} \rangle, \quad \langle \mathcal{O}^\dagger \mathcal{O} \rangle \geq 0
    \end{split}
\end{equation}
We can impose these conditions for $\mathcal{O} = x^m y^n p_x^s p_y^t$, with $\ell(\mathcal{O})=m+n+2s+2t$ smaller than a cutoff. It turns out that the ``free variables" in this case are those of the form $\langle x^m \rangle$, and all other variables can be solved as linear functions~\footnote{The coefficients depend on $E$, the energy of the eigenstate.} of $\langle x^m \rangle$. This results in a linear growth of the number of free variables, in sharp contrast to the exponential growth shown in Table~\ref{tab: freevariables} for the BFSS model in the main text.

Next, we scan over different values of $E$ and impose the positivity condition (similar to the method used in the main text), producing Figure~\ref{fig: x2y2} at a level cutoff of $12$. At this level, the highest moment is $\langle x^{12} \rangle$. The result is clear: there are only several islands of allowed regions for the eigenvalue $E$, each corresponding to an eigenstate of the Hamiltonian in Eq.~\eqref{eq: toyH}. 

Moreover, our bootstrap results show evidence that the spectrum of the Hamiltonian in Eq.~\eqref{eq: toyH} is discrete, a fact originally proven in the last century~\cite{Hoppe1980, Simon:1983jy}.
\subsection{Ground state bootstrap}

\begin{figure}[t]
\centering
\includegraphics[width=0.70\textwidth]{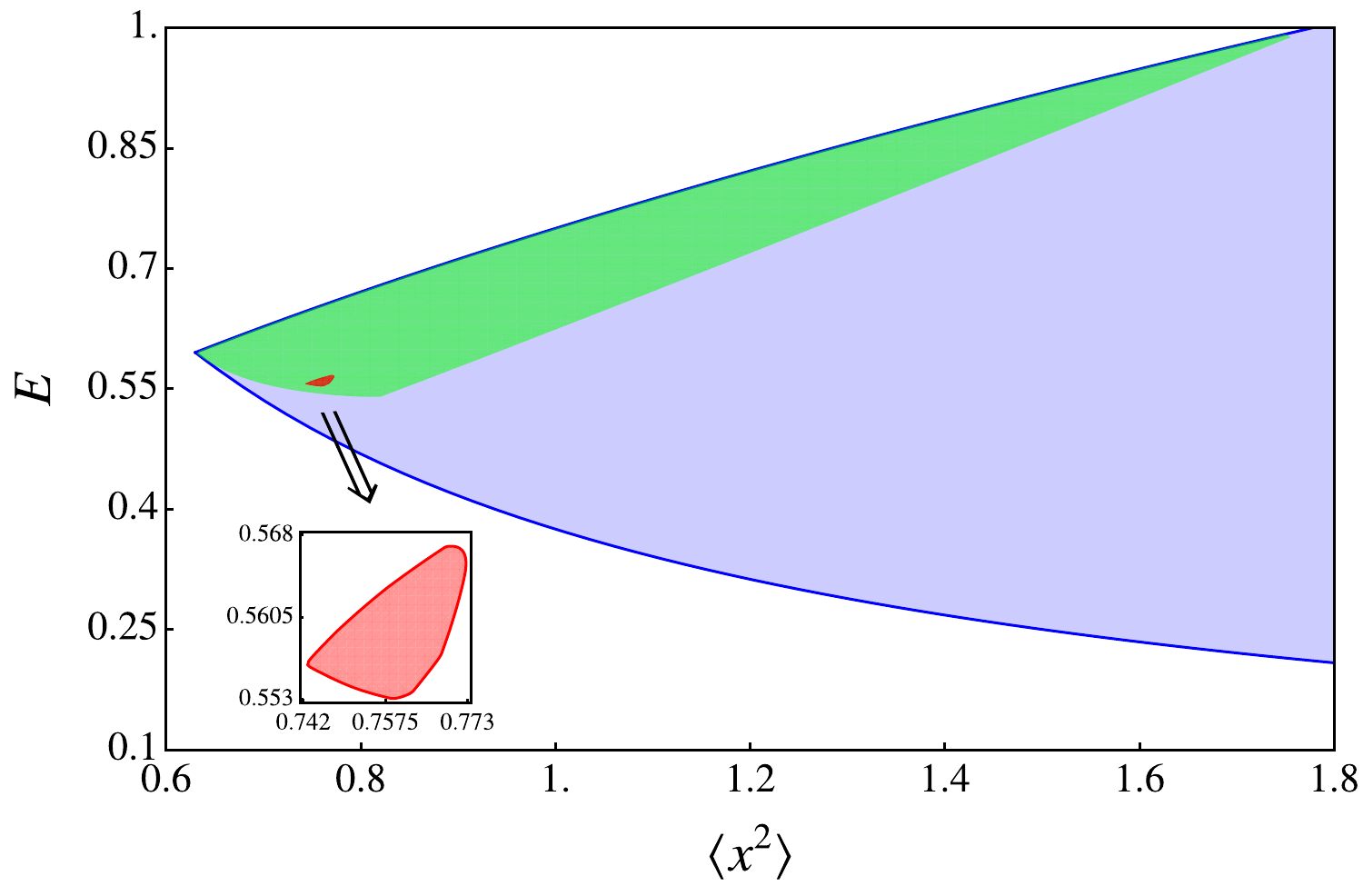}

\caption{\label{fig: toyground}{Bootstrap bounds on the ground state energy and $\langle x^2 \rangle$. Here, blue, green, and red correspond to level cutoffs of ${\color{blue} 4}$, ${\color{dgreen} 6}$, and ${\color{red} 8}$, respectively. The shaded regions represent the allowed regions; at higher levels, the peninsula shrinks to a {\color{dgreen} large island} and eventually a {\color{red} tiny island}. Note also that the {\red tiny island} is still relatively close to the tip of the {\color{blue} peninsula}. }
}
\end{figure}

The result in Appendix~\ref{sec: oldschool} is deceptively promising and does not serve as a good source of intuition for the convergence behavior of the BFSS model discussed in the main text. The reason is that, for the BFSS model, the following equation becomes trivial due to large-$N$ factorization:
\begin{equation}
    \langle \mathcal{O} H \rangle = \langle H \mathcal{O} \rangle = E \langle \mathcal{O} \rangle
\end{equation}
Instead, our equivalent condition here would be:
\begin{equation}
    \langle [H, \mathcal{O}] \rangle = 0
\end{equation}
which holds for any stationary state, including the canonical thermal ensemble, etc.\footnote{This equation is much weaker, leading to a power-law growth in the number of free variables as the level cutoff increases.}. We can select the ground state using the positivity condition discussed in Appendix~\ref{app: gpositivity}:
\begin{equation}
    \langle \mathcal{O}^\dagger [H, \mathcal{O}] \rangle \geq 0, \quad \langle \mathcal{O}^\dagger \mathcal{O} \rangle \geq 0
\end{equation}

The result of the bootstrap is summarized in Figure~\ref{fig: toyground}. Initially, at level $4$, we do not have a compact allowed region, but it eventually shrinks to a small island. This gives us hope that the allowed region for the BFSS model discussed in the main text (Figure~\ref{fig: plconvergence}) will ultimately shrink to a compact region.
\section{Invariant Tensors}

As discussed in Section~\ref{sec: variables}, we must choose a basis for invariant tensors. This is rather straightforward for bosonic irreps but somewhat subtle for fermionic irreps. In this appendix, we provide more details on this topic.

\subsection{Purely Bosonic}\label{app: purebos}

The situation concerning invariant tensors with only vector indices is straightforward, and we can define a canonical basis. For fewer than $9$ vector indices, the number of invariants corresponds to the number of Wick contractions between different indices\footnote{However, for the number of indices equal to or greater than $9$, this is not true. For example, at the 9-index level, we have $\epsilon^{I_1 \cdots I_9}$.}. %
For example,
\begin{small}
\begin{equation}
\begin{split}
    \mathcal{I}^{IJ}&: \, \delta^{IJ}, \\
    \mathcal{I}^{IJKL}&: \,\delta^{IJ}\delta^{KL},\, \delta^{IK}\delta^{JL},\, \delta^{IL}\delta^{JK}, \\
    \mathcal{I}^{IJKLMN}&: \delta^{IN}\delta^{JM}\delta^{KL}, \delta^{IM}\delta^{JN}\delta^{KL}, \delta^{IN}\delta^{JL}\delta^{KM}, \delta^{IL}\delta^{JN}\delta^{KM}, \delta^{IM}\delta^{JL}\delta^{KN}, \\
    & \,\,\,\,\,\delta^{IL}\delta^{JM}\delta^{KN}, \delta^{IN}\delta^{JK}\delta^{LM}, \delta^{IK}\delta^{JN}\delta^{LM}, \delta^{IJ}\delta^{KN}\delta^{LM}, \delta^{IM}\delta^{JK}\delta^{LN}, \\
    & \,\,\,\,\, \delta^{IK}\delta^{JM}\delta^{LN}, \delta^{IJ}\delta^{KM}\delta^{LN}, \delta^{IL}\delta^{JK}\delta^{MN}, \delta^{IK}\delta^{JL}\delta^{MN}, \delta^{IJ}\delta^{KL}\delta^{MN}, \\
    & \cdots
\end{split}
\end{equation}
\end{small}

The dimension of the space of invariant tensors is fixed by the decomposition of the following tensor product representations:
\begin{equation}\label{eq: counting}
    \begin{split}
        {\bf 9} \times \mathbf{9}  &= \mathbf{1} + \mathbf{36} + \mathbf{44}, \\
        {\bf 9} \times \mathbf{9} \times \mathbf{9} \times \mathbf{9} &= 3 (\mathbf{1}) + \cdots, \\
        {\bf 9} \times \mathbf{9} \times \mathbf{9} \times \mathbf{9} \times \mathbf{9} \times \mathbf{9} &= 15 (\mathbf{1}) + \cdots.
    \end{split}
\end{equation}

Compared to the situation involving the $\gamma$ matrices, which we will discuss shortly, the purely bosonic case is greatly simplified: reshuffling the vector indices only leads to a permutation of the basis. As a result, the ordering of the vector indices is not critical here. Another consequence of this is that the list of bases for pure bosonic invariant tensors can be generated recursively.

\subsection{Gamma algebra \label{app: GammaAlgebra}}
Before diving into the discussion of invariant tensors corresponding to both the vector and spinor representations, we first summarize some formulas for the gamma matrices and establish the conventions used throughout this work\footnote{More explicitly, we are using the \textit{friendly representation} in \cite{Freedman:2012zz}, where all the gamma matrix elements are real.}.

The gamma matrices are defined to satisfy\footnote{We could treat $\delta_{\alpha \beta}$ as the gamma matrix with zero indices.}:
\begin{align}
    \{ \gamma^I, \gamma^J \} = 2 \delta^{IJ}, \quad \gamma^{I_1 I_2 \cdots I_n} = \gamma^{[I_1} \gamma^{I_2} \cdots \gamma^{I_n]}
\end{align}
It turns out that gamma matrices with up to four indices already form a linear basis of gamma matrices, and those with more indices satisfy:
\begin{equation}\label{eq: mirror1}
    \gamma^{\text{ind}^c} = \frac{1}{\text{Length}[\text{ind}]!}\epsilon^{\text{ind}^c\,\text{ind}^r}\gamma^{\text{ind}}
\end{equation}
Here $\text{ind}^c$ denotes the complement of indices from $1$ to $9$, and $\text{ind}^r$ denotes the reversal of the indices $\text{ind}$. For example, if the length of $\text{ind}$ is $l$, then the length of $\text{ind}^c$ is $9-l$. Here are some instant results of this equation:
\begin{equation}
    \gamma _{\alpha \beta }^{IJKLM} \gamma _{\eta \epsilon }^{JKLM} = \frac{1}{24!}\epsilon^{IJKLM M'L'K'J'}\gamma _{\alpha \beta }^{J'K'L'M'}\gamma _{\eta \epsilon }^{JKLM} = \gamma _{\alpha \beta }^{JKLM} \gamma _{\eta \epsilon }^{IJKLM}
\end{equation}
\begin{equation}
    \gamma^{IJKLM NOPQ}_{\alpha\beta} = \epsilon^{IJKLM NOPQ}\delta_{\alpha\beta}
\end{equation}

To process invariant tensors involving gamma matrices, we need to study the gamma bi-product in the following form:
\begin{equation}
    \gamma^{\texttt{open}_1 \sim \texttt{closed}}_{\alpha\beta} \gamma^{\texttt{open}_2 \sim \texttt{closed}}_{\eta\epsilon}
\end{equation}
Here $\texttt{closed}$ represents the bosonic indices summed over, while $\texttt{open}$ represents the indices indicating the representation of the bi-product. The length of $\texttt{closed}$ is what matters for the closed indices. The symbol $\sim$ means to join the two strings together.

We first introduce the sign factor for the gamma matrix in the friendly representation:
\begin{equation}
    \gamma^{\texttt{ind}}_{\alpha\beta} = \text{sgn}_\gamma(\ell) \gamma^{\texttt{ ind}}_{\beta\alpha}
\end{equation}
Here $\ell$ is the length of the bosonic indices. As $\ell$ varies from $0$ to $9$, the sign factor $\text{sgn}_\gamma(\ell)$ is given by:
\begin{equation}
   \text{sgn}_\gamma(\ell) = \{ 1,\, 1,\, -1,\, -1,\, 1,\, 1,\, -1,\, -1,\, 1,\, 1 \}.
\end{equation}

Just as Eq.~\eqref{eq: mirror1} applies for the gamma matrix, we have a similar mirror formula for the gamma bi-product:
\begin{equation}\label{eq: bi-product}
    \gamma^{\texttt{open}_1 \sim \texttt{closed}}_{\alpha\beta} \gamma^{\texttt{open}_2 \sim \texttt{closed}}_{\eta\epsilon} = f(\ell_1, \ell_2, c)\gamma_{\alpha\beta}^{\texttt{open}_2 \sim \texttt{closed}^{'}} \gamma_{\eta\epsilon}^{\texttt{open}_1 \sim \texttt{closed}^{'}} + (\delta \text{ terms})
\end{equation}
Here $\ell_1$ and $\ell_2$ are the lengths of open index $1$ and open index $2$, respectively, and $c$ is the length of the closed index. The length of the closed index on the right-hand side is $9 - \ell_1 - \ell_2 - c$. The $(\delta \text{ terms})$ refer to terms involving $\delta$ with open indices, which are non-zero only when we have more than one open index in the bi-product. The additional factor $f$ reads:
\begin{equation}
    f(\ell_1, \ell_2, c) = \text{sgn}_\gamma(\ell_1 + \ell_2) \times \frac{c!}{(9 - \ell_1 - \ell_2 - c)!}
\end{equation}

As an example, we have:
\begin{equation}
    \gamma^{I \sim \texttt{closed}}_{\alpha\beta} \gamma^{J \sim \texttt{closed}}_{\eta\epsilon} = f(\ell_1, \ell_2, c)\gamma_{\alpha\beta}^{J \sim \texttt{closed}^{'}} \gamma_{\eta\epsilon}^{I \sim \texttt{closed}^{'}} + \frac{1}{c+1} \delta^{IJ}\gamma^{K \sim \texttt{closed}}_{\alpha\beta} \gamma^{K \sim \texttt{closed}}_{\eta\epsilon}
\end{equation}
\subsection{Mixed Invariant tensors}
In this part of the appendix, we present some examples of our choice of invariant tensors when spinor indices are involved. For two spinor indices, we use the basis: %
\begin{small}
\begin{equation}
\begin{split}
    \mathcal{I}_{\beta\alpha}&: \, \delta _{\beta \alpha }, \\
    \mathcal{I}^I_{\beta\alpha}&: {\gamma _{\beta \alpha }^{I}}, \\
    \mathcal{I}^{IJ}_{\beta\alpha}&: {\delta _{\beta \alpha }\delta^{IJ}, \gamma _{\beta \alpha } ^{IJ}}, \\
    \mathcal{I}^{IJK}_{\beta\alpha}&: {\gamma _{\beta \alpha }^{IJK}, \gamma _{\beta \alpha }^{K}\delta ^{IJ}, \gamma _{\beta \alpha }^{J}\delta ^{IK}, \gamma _{\beta \alpha }^{I}\delta ^{JK}}, \\
    & \cdots
\end{split}
\end{equation}
\end{small}
For four spinor indices, we choose %
\begin{small}
\begin{equation}
\begin{split}
    \mathcal{I}_{\eta \epsilon\beta\alpha}&: \, {\delta _{\eta \epsilon }\delta _{\beta \alpha }, \gamma _{\eta \epsilon }^{I}\gamma _{\beta \alpha }^{I}, \gamma _{\eta \epsilon }^{IJ}\gamma _{\beta \alpha }^{IJ}, \gamma _{\eta \epsilon }^{IJK}\gamma _{\beta \alpha }^{IJK}, \gamma _{\eta \epsilon }^{IJKL}\gamma _{\beta \alpha }^{IJKL}}, \\
    \mathcal{I}^I_{\eta \epsilon\beta\alpha}&: \,{\gamma _{\eta \epsilon }^{I}\delta _{\beta \alpha }, \delta _{\eta \epsilon }\gamma _{\beta \alpha }^{I}, \gamma _{\eta \epsilon }^{IJ}\gamma _{\beta \alpha }^{J}, \gamma _{\eta \epsilon }^{J}\gamma _{\beta \alpha }^{IJ}, \gamma _{\eta \epsilon }^{IJK}\gamma _{\beta \alpha }^{JK}, \gamma _{\eta \epsilon }^{JK}\gamma _{\beta \alpha }^{IJK}, \gamma _{\eta \epsilon }^{IJKL}\gamma _{\beta \alpha }^{JKL}, \gamma _{\eta \epsilon }^{JKL}\gamma _{\beta \alpha }^{IJKL}, \gamma _{\eta \epsilon }^{IJKLM}\gamma _{\beta \alpha }^{JKLM}}, \\
    \mathcal{I}^{IJ}_{\eta \epsilon\beta\alpha}&: \gamma _{\eta \epsilon }^{IJ}\delta _{\beta \alpha }, \gamma _{\eta \epsilon }^{J}\gamma _{\beta \alpha }^{I}, \gamma _{\eta \epsilon }^{I}\gamma _{\beta \alpha }^{J}, \gamma _{\eta \epsilon }^{IJK}\gamma _{\beta \alpha }^{K}, \delta _{\eta \epsilon } \gamma _{\beta \alpha }^{IJ}, \gamma _{\eta \epsilon }^{JK}\gamma _{\beta \alpha }^{IK}, \gamma _{\eta \epsilon }^{IK}\gamma _{\beta \alpha }^{JK}, \gamma _{\eta \epsilon }^{IJKL}\gamma _{\beta \alpha }^{KL}, \gamma _{\eta \epsilon }^{K}\gamma _{\beta \alpha }^{IJK}, \gamma _{\eta \epsilon }^{JKL}\gamma _{\beta \alpha }^{IKL},\\
    &\quad \gamma _{\eta \epsilon }^{IKL}\gamma _{\beta \alpha }^{JKL}, \gamma _{\eta \epsilon }^{IJKLM}\gamma _{\beta \alpha }^{KLM}, \gamma _{\eta \epsilon }^{KL}\gamma _{\beta \alpha }^{IJKL}, \gamma _{\eta \epsilon }^{JKLM}\gamma _{\beta \alpha }^{IKLM}, \gamma _{\eta \epsilon }^{IKLM}\gamma _{\beta \alpha }^{JKLM}, \gamma _{\eta \epsilon }^{KLM}\gamma _{\beta \alpha }^{IJKLM}, \beta _{\eta \epsilon }\delta _{\beta \alpha }\delta ^{IJ},\\
    &\quad \gamma _{\eta \epsilon }^{K}\gamma _{\beta \alpha }^{K}\delta ^{IJ}, \gamma _{\eta \epsilon }^{KL}\gamma _{\beta \alpha }^{KL}\delta ^{IJ}, \gamma _{\eta \epsilon }^{KLM}\gamma _{\beta \alpha }^{KLM}\delta ^{IJ}, \gamma _{\eta \epsilon }^{KLMN}\gamma _{\beta \alpha }^{KLMN}\delta ^{IJ}, \\
    & \cdots
\end{split}
\end{equation}
\end{small}
And finally, for the case with six spinor indices (a total of 55 invariant tensors):
\begin{small}
\begin{equation}
\begin{split}
    \mathcal{I}_{\mu \lambda \eta \epsilon \beta\alpha}&: \, \delta _{\mu \lambda }\delta _{\eta \epsilon }\delta _{\beta \alpha }, \delta _{\mu \lambda }\gamma _{\eta \epsilon }^{I}\gamma _{\beta \alpha }^{I}, \delta _{\mu \lambda }\gamma _{\eta \epsilon }^{IJ}\gamma _{\beta \alpha }^{IJ}, \delta _{\mu \lambda }\gamma _{\eta \epsilon }^{IJK}\gamma _{\beta \alpha }^{IJK}, \delta _{\mu \lambda }\gamma _{\eta \epsilon }^{IJKL}\gamma _{\beta \alpha }^{IJKL}, \gamma _{\mu \lambda }^{I}\gamma _{\eta \epsilon }^{I}\delta _{\beta \alpha }, \gamma _{\mu \lambda }^{I}\delta _{\eta \epsilon }\gamma _{\beta \alpha }^{I}, \gamma _{\mu \lambda }^{I}\gamma _{\eta \epsilon }^{IJ}\gamma _{\beta \alpha }^{J},\\
    &\quad \gamma _{\mu \lambda }^{I}\gamma _{\eta \epsilon }^{J}\gamma _{\beta \alpha }^{IJ}, \gamma _{\mu \lambda }^{I}\gamma _{\eta \epsilon }^{IJK}\gamma _{\beta \alpha }^{JK}, \gamma _{\mu \lambda }^{I}\gamma _{\eta \epsilon }^{JK}\gamma _{\beta \alpha }^{IJK}, \gamma _{\mu \lambda }^{I}\gamma _{\eta \epsilon }^{IJKL}\gamma _{\beta \alpha }^{JKL}, \gamma _{\mu \lambda }^{I}\gamma _{\eta \epsilon }^{JKL}\gamma _{\beta \alpha }^{IJKL}, \gamma _{\mu \lambda }^{I}\gamma _{\eta \epsilon }^{IJKLM}\gamma _{\beta \alpha }^{JKLM},\\
    &\quad \gamma _{\mu \lambda }^{IJ}\gamma _{\eta \epsilon }^{IJ}\delta _{\beta \alpha }, \gamma _{\mu \lambda }^{IJ}\gamma _{\eta \epsilon }^{I}\gamma _{\beta \alpha }^{J}, \gamma _{\mu \lambda }^{IJ}\gamma _{\eta \epsilon }^{IJK}\gamma _{\beta \alpha }^{K}, \gamma _{\mu \lambda }^{IJ}\delta _{\eta \epsilon }\gamma _{\beta \alpha }^{IJ}, \gamma _{\mu \lambda }^{IJ}\gamma _{\eta \epsilon }^{IK}\gamma _{\beta \alpha }^{JK}, \gamma _{\mu \lambda }^{IJ}\gamma _{\eta \epsilon }^{IJKL}\gamma _{\beta \alpha }^{KL}, \gamma _{\mu \lambda }^{IJ}\gamma _{\eta \epsilon }^{K}\gamma _{\beta \alpha }^{IJK}, \gamma _{\mu \lambda }^{IJ}\gamma _{\eta \epsilon }^{IKL}\gamma _{\beta \alpha }^{JKL},\\
    &\quad \gamma _{\mu \lambda }^{IJ}\gamma _{\eta \epsilon }^{IJKLM}\gamma _{\beta \alpha }^{KLM}, \gamma _{\mu \lambda }^{IJ}\gamma _{\eta \epsilon }^{KL}\gamma _{\beta \alpha }^{IJKL}, \gamma _{\mu \lambda }^{IJ}\gamma _{\eta \epsilon }^{IKLM}\gamma _{\beta \alpha }^{JKLM}, \gamma _{\mu \lambda }^{IJ}\gamma _{\eta \epsilon }^{KLM}\gamma _{\beta \alpha }^{IJKLM}, \gamma _{\mu \lambda }^{IJK}\gamma _{\eta \epsilon }^{IJK}\delta _{\beta \alpha }, \gamma _{\mu \lambda }^{IJK}\gamma _{\eta \epsilon }^{IJ}\gamma _{\beta \alpha }^{K},\\
    &\quad \gamma _{\mu \lambda }^{IJK}\gamma _{\eta \epsilon }^{IJKL}\gamma _{\beta \alpha }^{L}, \gamma _{\mu \lambda }^{IJK}\gamma _{\eta \epsilon }^{I}\gamma _{\beta \alpha }^{JK}, \gamma _{\mu \lambda }^{IJK}\gamma _{\eta \epsilon }^{IJL}\gamma _{\beta \alpha }^{KL}, \gamma _{\mu \lambda }^{IJK}\gamma _{\eta \epsilon }^{IJKLM}\gamma _{\beta \alpha }^{LM}, \gamma _{\mu \lambda }^{IJK}\delta _{\eta \epsilon }\gamma _{\beta \alpha }^{IJK}, \gamma _{\mu \lambda }^{IJK}\gamma _{\eta \epsilon }^{IL}\gamma _{\beta \alpha }^{JKL},\\
    &\quad \gamma _{\mu \lambda }^{IJK}\gamma _{\eta \epsilon }^{IJLM}\gamma _{\beta \alpha }^{KLM}, \gamma _{\mu \lambda }^{IJK}\gamma _{\eta \epsilon }^{IJKLMN}\gamma _{\beta \alpha }^{LMN}, \gamma _{\mu \lambda }^{IJK}\gamma _{\eta \epsilon }^{L}\gamma _{\beta \alpha }^{IJKL}, \gamma _{\mu \lambda }^{IJK}\gamma _{\eta \epsilon }^{ILM}\gamma _{\beta \alpha }^{JKLM}, \gamma _{\mu \lambda }^{IJK}\gamma _{\eta \epsilon }^{IJLMN}\gamma _{\beta \alpha }^{KLMN},\\
    &\quad \gamma _{\mu \lambda }^{IJK}\gamma _{\eta \epsilon }^{LM}\gamma _{\beta \alpha }^{IJKLM}, \gamma _{\mu \lambda }^{IJKL}\gamma _{\eta \epsilon }^{IJKL}\delta _{\beta \alpha }, \gamma _{\mu \lambda }^{IJKL}\gamma _{\eta \epsilon }^{IJK}\gamma _{\beta \alpha }^{L}, \gamma _{\mu \lambda }^{IJKL}\gamma _{\eta \epsilon }^{IJKLM}\gamma _{\beta \alpha }^{M}, \gamma _{\mu \lambda }^{IJKL}\gamma _{\eta \epsilon }^{IJ}\gamma _{\beta \alpha }^{KL}, \gamma _{\mu \lambda }^{IJKL}\gamma _{\eta \epsilon }^{IJKM}\gamma _{\beta \alpha }^{LM},\\
    &\quad \gamma _{\mu \lambda }^{IJKL}\gamma _{\eta \epsilon }^{IJKLMN}\gamma _{\beta \alpha }^{MN}, \gamma _{\mu \lambda }^{IJKL}\gamma _{\eta \epsilon }^{I}\gamma _{\beta \alpha }^{JKL}, \gamma _{\mu \lambda }^{IJKL}\gamma _{\eta \epsilon }^{IJM}\gamma _{\beta \alpha }^{KLM}, \gamma _{\mu \lambda }^{IJKL}\gamma _{\eta \epsilon }^{IJKMN}\gamma _{\beta \alpha }^{LMN}, \gamma _{\mu \lambda }^{IJKL}\delta _{\eta \epsilon }\gamma _{\beta \alpha }^{IJKL},\\
    &\quad \gamma _{\mu \lambda }^{IJKL}\gamma _{\eta \epsilon }^{IM}\gamma _{\beta \alpha }^{JKLM}, \gamma _{\mu \lambda }^{IJKL}\gamma _{\eta \epsilon }^{IJMN}\gamma _{\beta \alpha }^{KLMN}, \gamma _{\mu \lambda }^{IJKL}\gamma _{\eta \epsilon }^{M}\gamma _{\beta \alpha }^{IJKLM}, \gamma _{\mu \lambda }^{IJKL}\gamma _{\eta \epsilon }^{IMN}\gamma _{\beta \alpha }^{JKLMN}, \gamma _{\mu \lambda }^{IJKL}\gamma _{\eta \epsilon }^{MN}\gamma _{\beta \alpha }^{IJKLMN}, 
\end{split}
\end{equation}
\end{small}
We note that the number of tensors in the basis always matches the multiplicity of singlets in the tensor product decomposition. Parallel to Eq.~\eqref{eq: counting}, we have, for example:
\begin{equation}
    \mathbf{16} \times \mathbf{16} \times \mathbf{16} \times \mathbf{16} \times \mathbf{16} \times \mathbf{16} = 55 (\mathbf{1}) + \cdots
\end{equation}
An important distinction from the pure bosonic tensors discussed in Appendix~\ref{app: purebos} is that the ordering of spinor indices is crucial. This is exemplified by the crossing relation, as discussed in Section~\ref{sec: example}:
\begin{equation}
    \begin{split}
        \delta_{\eta \alpha}\delta_{\beta \epsilon}&=\frac{1}{16} \gamma _{\beta \alpha }^I \gamma _{\eta \epsilon }^I+\frac{1}{16} \delta _{\beta \alpha } \delta _{\eta \epsilon }-\frac{1}{32} \gamma _{\beta \alpha }^{IJ} \gamma _{\eta \epsilon }^{IJ}-\frac{1}{96} \gamma _{\beta \alpha }^{IJK} \gamma _{\eta \epsilon }^{IJK}+\frac{1}{384} \gamma _{\beta \alpha }^{IJKL} \gamma _{\eta \epsilon }^{IJKL}\\
        \gamma_{\eta \alpha}^I\gamma_{\beta \epsilon}^I&=-\frac{7}{16} \gamma _{\beta \alpha }^I \gamma _{\eta \epsilon }^I+\frac{9}{16} \delta _{\beta \alpha } \delta _{\eta \epsilon }-\frac{5}{32} \gamma _{\beta \alpha }^{IJ} \gamma _{\eta \epsilon }^{IJ}+\frac{1}{32} \gamma _{\beta \alpha }^{IJK} \gamma _{\eta \epsilon }^{IJK}+\frac{1}{384} \gamma _{\beta \alpha }^{IJKL} \gamma _{\eta \epsilon }^{IJKL}\\
        \gamma_{\eta \alpha}^{IJ} \gamma_{\beta \epsilon}^{IJ}&=-\frac{5}{2} \gamma _{\beta \alpha }^I \gamma _{\eta \epsilon }^I-\frac{9}{2} \delta _{\beta \alpha } \delta _{\eta \epsilon }+\frac{1}{2} \gamma _{\beta \alpha }^{IJ} \gamma _{\eta \epsilon }^{IJ}+\frac{1}{48} \gamma _{\beta \alpha }^{IJKL} \gamma _{\eta \epsilon }^{IJKL}\\
        \gamma_{\eta \alpha}^{IJK} \gamma_{\beta \epsilon}^{IJK}&=\frac{21}{2} \gamma _{\beta \alpha }^I \gamma _{\eta \epsilon }^I-\frac{63}{2} \delta _{\beta \alpha } \delta _{\eta \epsilon }+\frac{1}{2} \gamma _{\beta \alpha }^{IJK} \gamma _{\eta \epsilon }^{IJK}+\frac{1}{16} \gamma _{\beta \alpha }^{IJKL} \gamma _{\eta \epsilon }^{IJKL}\\
        \gamma_{\eta \alpha}^{IJKL} \gamma_{\beta \epsilon}^{IJKL}&=21 \gamma _{\beta \alpha }^I \gamma _{\eta \epsilon }^I+189 \delta _{\beta \alpha } \delta _{\eta \epsilon }+\frac{21}{2} \gamma _{\beta \alpha }^{IJ} \gamma _{\eta \epsilon }^{IJ}+\frac{3}{2} \gamma _{\beta \alpha }^{IJK} \gamma _{\eta \epsilon }^{IJK}+\frac{3}{8} \gamma _{\beta \alpha }^{IJKL} \gamma _{\eta \epsilon }^{IJKL}.
    \end{split}
\end{equation}

\section{Extrapolation of Monte Carlo data \label{extrapolate}}
Here we discuss some details of the extrapolation of the Monte Carlo data \cite{Berkowitz:2016jlq}.\footnote{We thank Masanori Hanada and Stratos Pateloudis for some discussions on the data.} We used the results reported in the Appendix B of \cite{Berkowitz:2016jlq}. To compare to our results, we in principle need to perform a 3-variable extrapolation $(1/L, 1/N, T) \to 0$, where $L$ is the number of lattice sites.  In practice, we find that the statistical error on the Monte Carlo measurements is negligible compared to the potential systematic error due to extrapolation. 
For the correlator $\ev{\tr[X^I,X^J]^2}$ we found a good fit using the ansatz
\begin{align}
\ev{\tr[X^I,X^J]^2} =  b +  a_T T^{9/5} +  \frac{b_N}{N^2}+  \frac{b_L}{L},
\end{align}
where we chose the $T^{9/5}$ scaling motivated by the low-temperature scaling of energy with temperature. For the radius $\ev{\tr X^2}$, we did not find a particularly good fit. Indeed, the temperature dependence of $\ev{\tr X^2}$ reported in \cite{Berkowitz:2016jlq} is not monotonic, so we simply averaged over the temperature data from $T = 0.4$ to $T=0.7$ and used the simpler ansatz
\begin{align}
\ev{\tr X^2} =  a +  \frac{a_N}{N^2}+  \frac{a_L}{L}.
\end{align}
One could also compare to the most recent Monte Carlo simulations in \cite{Pateloudis:2022ijr}. These are performed at finite BMN \cite{Berenstein:2002jq} mass parameter  $\mu$. This stabilizes the BFSS model at finite $N$ by removing the flat directions. To compare to our results, we in principle need to perform a 4 parameter extrapolation $(1/L, 1/N, \mu, T) \to 0$, where $L$ is the number of lattice sites. For the energy, the deviation is about $1\%$ compared to the $\mu = 0$ value \cite{Pateloudis:2022ijr}. However, for $\ev{\tr X^2}$, the authors \cite{Pateloudis:2022ijr} observe a peculiar behavior at low temperatures $T=0.3$. In particular,  $\ev{\tr X^2}$ seems to initially increase as the mass parameter $\mu$ gets smaller. This likely indicates the difficulty of computing the observable $\ev{\tr X^2}$ which is highly sensitive to the potential decay of the metastable state. (Indeed there are vacua of the BMN model which have large $\ev{\tr X^2}$ at small $\mu$, which roughly speaking become scattering states as $\mu \to 0$ as opposed to the normalizable bound state.)

One can also consider the results reported in \cite{Berkowitz:2018qhn}. According to \cite{Maldacena:2018vsr, Berkowitz:2018qhn}, simple observables should agree in the low temperature, 't Hooft limit between the gauged and ungauged model. At low temperatures, \cite{Berkowitz:2018qhn} report an uncertainty in $\ev{\tr X^2}$ of a few percent, see their Figure 7.

\bibliography{ref}

\providecommand{\href}[2]{#2}\begingroup\raggedright\begin{thebibliography}{10}

\bibitem{deWit:1988wri}
B.~de~Wit, J.~Hoppe and H.~Nicolai, \emph{{On the Quantum Mechanics of Supermembranes}}, \href{https://doi.org/10.1201/9781482268737-11}{\emph{Nucl. Phys. B} {\bfseries 305} (1988) 545}.

\bibitem{Banks:1996vh}
T.~Banks, W.~Fischler, S.~H. Shenker and L.~Susskind, \emph{{M theory as a matrix model: A conjecture}}, \href{https://doi.org/10.1201/9781482268737-37}{\emph{Phys. Rev. D} {\bfseries 55} (1997) 5112} [\href{https://arxiv.org/abs/hep-th/9610043}{{\ttfamily hep-th/9610043}}].

\bibitem{Itzhaki:1998dd}
N.~Itzhaki, J.~M. Maldacena, J.~Sonnenschein and S.~Yankielowicz, \emph{{Supergravity and the large N limit of theories with sixteen supercharges}}, \href{https://doi.org/10.1103/PhysRevD.58.046004}{\emph{Phys. Rev. D} {\bfseries 58} (1998) 046004} [\href{https://arxiv.org/abs/hep-th/9802042}{{\ttfamily hep-th/9802042}}].

\bibitem{Maldacena:2023acv}
J.~Maldacena, \emph{{A simple quantum system that describes a black hole}},  \href{https://arxiv.org/abs/2303.11534}{{\ttfamily 2303.11534}}.

\bibitem{Kabat:2000zv}
D.~N. Kabat, G.~Lifschytz and D.~A. Lowe, \emph{{Black hole thermodynamics from calculations in strongly coupled gauge theory}}, \href{https://doi.org/10.1103/PhysRevLett.86.1426}{\emph{Int. J. Mod. Phys. A} {\bfseries 16} (2001) 856} [\href{https://arxiv.org/abs/hep-th/0007051}{{\ttfamily hep-th/0007051}}].

\bibitem{Anagnostopoulos:2007fw}
K.~N. Anagnostopoulos, M.~Hanada, J.~Nishimura and S.~Takeuchi, \emph{{Monte Carlo studies of supersymmetric matrix quantum mechanics with sixteen supercharges at finite temperature}}, \href{https://doi.org/10.1103/PhysRevLett.100.021601}{\emph{Phys. Rev. Lett.} {\bfseries 100} (2008) 021601} [\href{https://arxiv.org/abs/0707.4454}{{\ttfamily 0707.4454}}].

\bibitem{Hanada:2008ez}
M.~Hanada, Y.~Hyakutake, J.~Nishimura and S.~Takeuchi, \emph{{Higher derivative corrections to black hole thermodynamics from supersymmetric matrix quantum mechanics}}, \href{https://doi.org/10.1103/PhysRevLett.102.191602}{\emph{Phys. Rev. Lett.} {\bfseries 102} (2009) 191602} [\href{https://arxiv.org/abs/0811.3102}{{\ttfamily 0811.3102}}].

\bibitem{Catterall:2008yz}
S.~Catterall and T.~Wiseman, \emph{{Black hole thermodynamics from simulations of lattice Yang-Mills theory}}, \href{https://doi.org/10.1103/PhysRevD.78.041502}{\emph{Phys. Rev. D} {\bfseries 78} (2008) 041502} [\href{https://arxiv.org/abs/0803.4273}{{\ttfamily 0803.4273}}].

\bibitem{Filev:2015hia}
V.~G. Filev and D.~O'Connor, \emph{{The BFSS model on the lattice}}, \href{https://doi.org/10.1007/JHEP05(2016)167}{\emph{JHEP} {\bfseries 05} (2016) 167} [\href{https://arxiv.org/abs/1506.01366}{{\ttfamily 1506.01366}}].

\bibitem{Kadoh:2015mka}
D.~Kadoh and S.~Kamata, \emph{{Gauge/gravity duality and lattice simulations of one dimensional SYM with sixteen supercharges}},  \href{https://arxiv.org/abs/1503.08499}{{\ttfamily 1503.08499}}.

\bibitem{Berkowitz:2016jlq}
E.~Berkowitz, E.~Rinaldi, M.~Hanada, G.~Ishiki, S.~Shimasaki and P.~Vranas, \emph{{Precision lattice test of the gauge/gravity duality at large-$N$}}, \href{https://doi.org/10.1103/PhysRevD.94.094501}{\emph{Phys. Rev. D} {\bfseries 94} (2016) 094501} [\href{https://arxiv.org/abs/1606.04951}{{\ttfamily 1606.04951}}].

\bibitem{Berkowitz:2018qhn}
E.~Berkowitz, M.~Hanada, E.~Rinaldi and P.~Vranas, \emph{{Gauged And Ungauged: A Nonperturbative Test}}, \href{https://doi.org/10.1007/JHEP06(2018)124}{\emph{JHEP} {\bfseries 06} (2018) 124} [\href{https://arxiv.org/abs/1802.02985}{{\ttfamily 1802.02985}}].

\bibitem{Pateloudis:2022ijr}
{\scshape Monte Carlo String/M-theory (MCSMC)} collaboration, \emph{{Precision test of gauge/gravity duality in D0-brane matrix model at low temperature}}, \href{https://doi.org/10.1007/JHEP03(2023)071}{\emph{JHEP} {\bfseries 03} (2023) 071} [\href{https://arxiv.org/abs/2210.04881}{{\ttfamily 2210.04881}}].

\bibitem{Catterall:2009xn}
S.~Catterall and T.~Wiseman, \emph{{Extracting black hole physics from the lattice}}, \href{https://doi.org/10.1007/JHEP04(2010)077}{\emph{JHEP} {\bfseries 04} (2010) 077} [\href{https://arxiv.org/abs/0909.4947}{{\ttfamily 0909.4947}}].

\bibitem{Anderson:2016rcw}
P.~D. Anderson and M.~Kruczenski, \emph{{Loop Equations and bootstrap methods in the lattice}}, \href{https://doi.org/10.1016/j.nuclphysb.2017.06.009}{\emph{Nucl. Phys. B} {\bfseries 921} (2017) 702} [\href{https://arxiv.org/abs/1612.08140}{{\ttfamily 1612.08140}}].

\bibitem{Lin:2020mme}
H.~W. Lin, \emph{{Bootstraps to strings: solving random matrix models with positivity}}, \href{https://doi.org/10.1007/JHEP06(2020)090}{\emph{JHEP} {\bfseries 06} (2020) 090} [\href{https://arxiv.org/abs/2002.08387}{{\ttfamily 2002.08387}}].

\bibitem{Han:2020bkb}
X.~Han, S.~A. Hartnoll and J.~Kruthoff, \emph{{Bootstrapping Matrix Quantum Mechanics}}, \href{https://doi.org/10.1103/PhysRevLett.125.041601}{\emph{Phys. Rev. Lett.} {\bfseries 125} (2020) 041601} [\href{https://arxiv.org/abs/2004.10212}{{\ttfamily 2004.10212}}].

\bibitem{Kazakov:2021lel}
V.~Kazakov and Z.~Zheng, \emph{{Analytic and numerical bootstrap for one-matrix model and \textquotedblleft{}unsolvable\textquotedblright{} two-matrix model}}, \href{https://doi.org/10.1007/JHEP06(2022)030}{\emph{JHEP} {\bfseries 06} (2022) 030} [\href{https://arxiv.org/abs/2108.04830}{{\ttfamily 2108.04830}}].

\bibitem{Kazakov:2022xuh}
V.~Kazakov and Z.~Zheng, \emph{{Bootstrap for lattice Yang-Mills theory}}, \href{https://doi.org/10.1103/PhysRevD.107.L051501}{\emph{Phys. Rev. D} {\bfseries 107} (2023) L051501} [\href{https://arxiv.org/abs/2203.11360}{{\ttfamily 2203.11360}}].

\bibitem{Cho:2022lcj}
M.~Cho, B.~Gabai, Y.-H. Lin, V.~A. Rodriguez, J.~Sandor and X.~Yin, \emph{{Bootstrapping the Ising Model on the Lattice}},  \href{https://arxiv.org/abs/2206.12538}{{\ttfamily 2206.12538}}.

\bibitem{Lin:2023owt}
H.~W. Lin, \emph{{Bootstrap bounds on D0-brane quantum mechanics}}, \href{https://doi.org/10.1007/JHEP06(2023)038}{\emph{JHEP} {\bfseries 06} (2023) 038} [\href{https://arxiv.org/abs/2302.04416}{{\ttfamily 2302.04416}}].

\bibitem{Kazakov:2024ool}
V.~Kazakov and Z.~Zheng, \emph{{Bootstrap for Finite N Lattice Yang-Mills Theory}},  \href{https://arxiv.org/abs/2404.16925}{{\ttfamily 2404.16925}}.

\bibitem{Li:2024wrd}
Z.~Li and S.~Zhou, \emph{{Bootstrapping the Abelian lattice gauge theories}}, \href{https://doi.org/10.1007/JHEP08(2024)154}{\emph{JHEP} {\bfseries 08} (2024) 154} [\href{https://arxiv.org/abs/2404.17071}{{\ttfamily 2404.17071}}].

\bibitem{Polchinski:1999br}
J.~Polchinski, \emph{{M theory and the light cone}}, \href{https://doi.org/10.1143/PTPS.134.158}{\emph{Prog. Theor. Phys. Suppl.} {\bfseries 134} (1999) 158} [\href{https://arxiv.org/abs/hep-th/9903165}{{\ttfamily hep-th/9903165}}].

\bibitem{LinZheng2}
H.~Lin and Z.~Zheng, \emph{Bootstrapping bfss, part ii}, {\emph{in prep} (2024) }.

\bibitem{Yi:1997eg}
P.~Yi, \emph{{Witten index and threshold bound states of D-branes}}, \href{https://doi.org/10.1016/S0550-3213(97)00486-0}{\emph{Nucl. Phys. B} {\bfseries 505} (1997) 307} [\href{https://arxiv.org/abs/hep-th/9704098}{{\ttfamily hep-th/9704098}}].

\bibitem{Moore:1998et}
G.~W. Moore, N.~Nekrasov and S.~Shatashvili, \emph{{D particle bound states and generalized instantons}}, \href{https://doi.org/10.1007/s002200050016}{\emph{Commun. Math. Phys.} {\bfseries 209} (2000) 77} [\href{https://arxiv.org/abs/hep-th/9803265}{{\ttfamily hep-th/9803265}}].

\bibitem{Konechny:1998vc}
A.~Konechny, \emph{{On asymptotic Hamiltonian for SU(N) matrix theory}}, \href{https://doi.org/10.1088/1126-6708/1998/10/018}{\emph{JHEP} {\bfseries 10} (1998) 018} [\href{https://arxiv.org/abs/hep-th/9805046}{{\ttfamily hep-th/9805046}}].

\bibitem{Porrati:1997ej}
M.~Porrati and A.~Rozenberg, \emph{{Bound states at threshold in supersymmetric quantum mechanics}}, \href{https://doi.org/10.1016/S0550-3213(97)00804-3}{\emph{Nucl. Phys. B} {\bfseries 515} (1998) 184} [\href{https://arxiv.org/abs/hep-th/9708119}{{\ttfamily hep-th/9708119}}].

\bibitem{Sethi:2000zf}
S.~Sethi and M.~Stern, \emph{{Invariance theorems for supersymmetric Yang-Mills theories}}, \href{https://doi.org/10.4310/ATMP.2000.v4.n2.a8}{\emph{Adv. Theor. Math. Phys.} {\bfseries 4} (2000) 487} [\href{https://arxiv.org/abs/hep-th/0001189}{{\ttfamily hep-th/0001189}}].

\bibitem{Gran:2001yh}
U.~Gran, \emph{{GAMMA: A Mathematica package for performing gamma matrix algebra and Fierz transformations in arbitrary dimensions}},  \href{https://arxiv.org/abs/hep-th/0105086}{{\ttfamily hep-th/0105086}}.

\bibitem{Feger:2019tvk}
R.~Feger, T.~W. Kephart and R.~J. Saskowski, \emph{{LieART 2.0 \textendash{} A Mathematica application for Lie Algebras and Representation Theory}}, \href{https://doi.org/10.1016/j.cpc.2020.107490}{\emph{Comput. Phys. Commun.} {\bfseries 257} (2020) 107490} [\href{https://arxiv.org/abs/1912.10969}{{\ttfamily 1912.10969}}].

\bibitem{Maldacena:2018vsr}
J.~Maldacena and A.~Milekhin, \emph{{To gauge or not to gauge?}}, \href{https://doi.org/10.1007/JHEP04(2018)084}{\emph{JHEP} {\bfseries 04} (2018) 084} [\href{https://arxiv.org/abs/1802.00428}{{\ttfamily 1802.00428}}].

\bibitem{Plefka:1997xq}
J.~Plefka and A.~Waldron, \emph{{On the quantum mechanics of M(atrix) theory}}, \href{https://doi.org/10.1016/S0550-3213(97)00762-1}{\emph{Nucl. Phys. B} {\bfseries 512} (1998) 460} [\href{https://arxiv.org/abs/hep-th/9710104}{{\ttfamily hep-th/9710104}}].

\bibitem{Frohlich:1999zf}
J.~Frohlich, G.~M. Graf, D.~Hasler, J.~Hoppe and S.-T. Yau, \emph{{Asymptotic form of zero energy wave functions in supersymmetric matrix models}}, \href{https://doi.org/10.1016/S0550-3213(99)00649-5}{\emph{Nucl. Phys. B} {\bfseries 567} (2000) 231} [\href{https://arxiv.org/abs/hep-th/9904182}{{\ttfamily hep-th/9904182}}].

\bibitem{Hoppe:2000tj}
J.~Hoppe and J.~Plefka, \emph{{The Asymptotic ground state of SU(3) matrix theory}},  \href{https://arxiv.org/abs/hep-th/0002107}{{\ttfamily hep-th/0002107}}.

\bibitem{Hasler:2002wt}
D.~Hasler and J.~Hoppe, \emph{{Asymptotic factorization of the ground state for SU(N) invariant supersymmetric matrix models}},  \href{https://arxiv.org/abs/hep-th/0206043}{{\ttfamily hep-th/0206043}}.

\bibitem{Lin:2014wka}
Y.-H. Lin and X.~Yin, \emph{{On the Ground State Wave Function of Matrix Theory}}, \href{https://doi.org/10.1007/JHEP11(2015)027}{\emph{JHEP} {\bfseries 11} (2015) 027} [\href{https://arxiv.org/abs/1402.0055}{{\ttfamily 1402.0055}}].

\bibitem{Berenstein:2002jq}
D.~E. Berenstein, J.~M. Maldacena and H.~S. Nastase, \emph{{Strings in flat space and pp waves from N=4 superYang-Mills}}, \href{https://doi.org/10.1088/1126-6708/2002/04/013}{\emph{JHEP} {\bfseries 04} (2002) 013} [\href{https://arxiv.org/abs/hep-th/0202021}{{\ttfamily hep-th/0202021}}].

\bibitem{Hanada:2021ipb}
M.~Hanada, \emph{{Bulk geometry in gauge/gravity duality and color degrees of freedom}}, \href{https://doi.org/10.1103/PhysRevD.103.106007}{\emph{Phys. Rev. D} {\bfseries 103} (2021) 106007} [\href{https://arxiv.org/abs/2102.08982}{{\ttfamily 2102.08982}}].

\bibitem{Fawzi:2023fpg}
H.~Fawzi, O.~Fawzi and S.~O. Scalet, \emph{{Certified algorithms for equilibrium states of local quantum Hamiltonians}}, \href{https://doi.org/10.1038/s41467-024-51592-3}{\emph{Nature Commun.} {\bfseries 15} (2024) 7394} [\href{https://arxiv.org/abs/2311.18706}{{\ttfamily 2311.18706}}].

\bibitem{Cho:2024kxn}
M.~Cho, B.~Gabai, J.~Sandor and X.~Yin, \emph{{Thermal Bootstrap of Matrix Quantum Mechanics}},  \href{https://arxiv.org/abs/2410.04262}{{\ttfamily 2410.04262}}.

\bibitem{deWit:1988xki}
B.~de~Wit, M.~Luscher and H.~Nicolai, \emph{{The Supermembrane Is Unstable}}, \href{https://doi.org/10.1016/0550-3213(89)90214-9}{\emph{Nucl. Phys. B} {\bfseries 320} (1989) 135}.

\bibitem{Balthazar:2016utu}
B.~Balthazar, V.~A. Rodriguez and X.~Yin, \emph{{Hamiltonian Truncation Study of Supersymmetric Quantum Mechanics: S-Matrix and Metastable States}}, \href{https://doi.org/10.1007/JHEP08(2019)100}{\emph{JHEP} {\bfseries 08} (2019) 100} [\href{https://arxiv.org/abs/1610.07275}{{\ttfamily 1610.07275}}].

\bibitem{Komatsu:2024vnb}
S.~Komatsu, A.~Martina, J.~a. Penedones, N.~Suchel, A.~Vuignier and X.~Zhao, \emph{{Gravity from quantum mechanics of finite matrices}},  \href{https://arxiv.org/abs/2401.16471}{{\ttfamily 2401.16471}}.

\bibitem{Hoppe1980}
J.~Hoppe, \emph{Two problems in quantum mechanics},  Master's thesis, Massachusetts Institute of Technology, 1980.

\bibitem{Simon:1983jy}
B.~Simon, \emph{{SOME QUANTUM OPERATORS WITH DISCRETE SPECTRUM BUT CLASSICALLY CONTINUOUS SPECTRUM}}, \href{https://doi.org/10.1016/0003-4916(83)90057-X}{\emph{Annals Phys.} {\bfseries 146} (1983) 209}.

\bibitem{Freedman:2012zz}
D.~Z. Freedman and A.~Van~Proeyen, \emph{{Supergravity}}. Cambridge Univ. Press, Cambridge, UK, 5, 2012, \href{https://doi.org/10.1017/CBO9781139026833}{10.1017/CBO9781139026833}.

\end{thebibliography}\endgroup
\bibliographystyle{jhep}

\end{document}